\documentclass{iopart}

\usepackage{iopams}
\usepackage{graphicx}

\begin{document}

\title[Virial expansion of molecular Brownian motion]{Virial
expansion of molecular Brownian motion \\
versus tales of ``statistical independency''}

\author{Yu E Kuzovlev}

\address{Donetsk Institute
for Physics and Technology of NASU, 83114 Donetsk, Ukraine}
\ead{kuzovlev@kinetic.ac.donetsk.ua}

\begin{abstract}
Basing on main principles of statistical mechanics only, an exact
virial expansion for path probability distribution of molecular
Brownian particle in a fluid is derived which connects response of
the distribution to perturbations of the fluid and statistical
correlations of its molecules with Brownian particle. The expansion
implies that (i) spatial spread of these correlations is finite, (ii)
this is inconsistent with Gaussian distribution involved by the
``molecular chaos'' hypothesis, and (iii) real path distribution
possesses power-law long tails. This means that actual Brownian path
never can be disjointed into statistically independent fragments,
even in the Boltzmann-Grad gas, but behaves as if Brownian particle's
diffusivity undergoes scaleless low-frequency fluctuations.
\end{abstract}


\vspace{2pc} \noindent{\it Keywords}\,: Rigorous results in
statistical mechanics, Kinetic theory of gases and liquids, Brownian
motion



\begin{verbatim}

                           ``God does not play dice''  ( A. Einstein )

\end{verbatim}

\section{Introduction}
Molecules in gases and liquids, or free electrons and holes in
crystals, etc., can be treated as ``small Brownian particles'' (BP)
since their thermal motion undoubtedly is ``not less random'' than
motion of the R.\,Brown's pollen suspended in water. Therefore,
seemingly, the A.\,Einstein's reasonings \cite{ae1,ae2} can be
applied to a small molecular-size BP too, again producing the
diffusion equation for probability density of BP's position. The
diffusion equation, in its turn, implies that probability
distribution, $\,V_0(t,\Delta{\bf R})\,$, of displacement of BP,
$\,\Delta {\bf R}\,$, during time interval $\,(0,t)\,$, at long
enough $\,t\,$ tends to the Gaussian distribution:
\begin{equation}
\begin{array}{c}
V_0(t,\Delta{\bf R})\,\rightarrow\,V_G(t,\Delta{\bf R})\,=\,(4\pi
Dt)^{-\,3/2}\,\exp{(-\Delta {\bf R}^2/4Dt)}\,\, \label{vg}
\end{array}
\end{equation}
Formally, Einstein in \cite{ae1} assumed that $\,\Delta {\bf R}\,$
consists of many  increments which are {\it statistically
independent} in the sense of the probability theory. Thus, from
statistical point of view, his result is equivalent to the ``\,{\it
law of large numbers}\,'' discovered by J.\,Bernoulli almost two
centuries earlier \cite{jb}.

The idea of {\it statistical independency}, in the form of the
``Stosszahlansatz'', or ``molecular chaos hypothesis'', was also
taken by L.\,Boltzmann as a principle of his molecular-kinetic theory
of gases \cite{bol}. Later in \cite{bog} N.\,Bogolyubov imparted it,
after non-principal modification, to a theory based on the
Bogolyubov-Born-Green-Kirkwood-Yvon (BBGKY) hierarchy of equations
\cite{bog,uf,re,bal,ll2}. It was violence against own brainchild,
because the BBGKY equations are quite all-sufficient ones and do not
need in any add-on (except natural initial conditions, of course).
Unfortunately, the violence lasts up to now\,: various (higher-order,
non-local, etc.) generalizations of the Boltzmann kinetic equation to
relatively dense gases \cite{uf,re,bal,ll2} all the same rest upon
one or another variant of the ``Stosszahlansatz''. Although an
attempt to break this tradition was made in \cite{i1} (see also
\cite{i2}).

At same time of the forties, N.\,Krylov in his book \cite{kr} (first
published in 1950 in Russian) argued that probability-theoretic
concepts under use in modern kinetics generally are incompatible with
principles of statistical mechanics. Especially he emphasized fallacy
of common opinions that ``\,{\it probabilities do exist regardless of
a theoretical construct and full-scale experiments}\,'' and that
``\,{\it obviously independent phenomena should possess independent
probability distributions}\,'' (italics mark brief citations from
\cite{kr}). According to Krylov, ``physical independency'' of events
in reality does not mean their ``{\it statistical independency}'' in
theory.

At present, survivability of the prejudices disclosed by Krylov is
the only excuse of the ``Stosszahlansatz'' or similar conjectures.
They gave rise to conviction that in the limit of ``Boltzmann-Grad
gas'' (``infinitely dilute gas'') infinite BBGKY hierarchy reduces to
the single Boltzmann equation, i.e. the latter presents exact gas
kinetics. If such was the case, then random walk of a molecular BP
(e.g. test or marked gas atom) would be made of many {\it
statistically independent} events and pieces. Then the ``{\it law of
large numbers}\,'' is in effect, and probability distribution of the
Brownian path $\,\Delta {\bf R}\,$ in thermodynamically equilibrium
gas should have Gaussian asymptotic (\ref{vg}).

But this is not true\,! In fact, as we will show, $\,V_0(t,\Delta{\bf
R})\,$ has essentially non-Gaussian asymptotic possessing power-law
tails at $\,\Delta {\bf R}^2/4Dt >1\,$ instead of the exponential
ones, even in equilibrium gas under the Boltzmann-Grad limit (BGL).

Hence, N.\,Krylov was right, and {\it statistical independency} of
colliding molecules or {\it statistical independency} of different
pieces of Brownian trajectory, etc., like {\it statistical
independency} of mathematical dice tosses, exists in imagination only
but not in physical reality. If reinterpreting the well-known words
said by Einstein, one can say that he was playing dice in \cite{ae1}
but God does not play dice.

A substantiation of these statements below (see also
\cite{pro,lp,p71}) will be done ``at very thermodynamical level''
basing on only determinism and reversibility of Hamiltonian
microscopic dynamics and besides general notions about many-particle
distribution functions and correlation functions \cite{uf,re,bal} and
main fluid properties.

In spite of so abstract approach, remarkably, our conclusions will be
in full qualitative agreement with result obtained in \cite{p1} by
means of crucial approximation of the BBGKY hierarchy and then its
direct solving under BGL:
\begin{equation}
\begin{array}{c}
V_0(t,\Delta {\bf R})\,\rightarrow \,(4\pi Dt)^{-\,3/2}\,\Gamma
(7/2)\, (1+\Delta {\bf R}^2/4Dt)^{-\,7/2}\,\label{ngl}
\end{array}
\end{equation}
(in \cite{p1} designation $\,W_1\,$ was used in place of $\,V_0\,$).
The diffusivity $\,D\,$ here, as well as in (\ref{vg}), is defined by
$\,\int \Delta {\bf R}^2\, V_0(t,\Delta {\bf R})\,d\Delta {\bf
R}\,\rightarrow \,6Dt\,$, while the arrow everywhere denotes
asymptotic at $\,t\,$ much greater than BP's mean free-flight time.

Our plan is as follows. In the beginning, a kind of virial expansion
for $\,V_0(t,\Delta{\bf R})\,$ is derived. It connects, from one
hand, first- and higher-order derivatives of $\,V_0(t,\Delta{\bf
R})\,$ with respect to gas density and, from the other hand, pair and
many-particle joint correlation functions (CF) of BP and gas. The CF
in their turn are definitely related to usual distribution functions
(DF). Since the latter by their sense are non-negative, the virial
expansion results in a series of differential inequalities to be
satisfied by $\,V_0(t,\Delta{\bf R})\,$. The first of them,
eventually, restricts a steepness of $\,V_0(t,\Delta{\bf R})$'s tails
and clearly forbids their exponentially fast decrease. At the end of
the paper, its relation to the problem of 1/f noise is commented.

\section{Virial expansion of Brownian path probability distribution}
\subsection{Full-scale experiments and fluctuation-dissipation relations}
We will consider a system consisting of $N\gg 1\,$ atoms in volume
$\,\Omega $ plus one more corpuscular ``Brownian particle'' (BP),
under the thermodynamical limit: $\,N\rightarrow\infty \,$,
$\,\Omega\rightarrow\infty \,$, $\,N/\Omega =\nu_{\,0}  =
\,$\,\,const\,. Initially, at time $\,t=0\,$, position of BP\,
$\,{\bf R}(t)\,$\, is definitely known. We are interested in the
already mentioned distribution $\,V_0(t,\Delta{\bf R})\,$ of
consequent BP's path $\,\Delta{\bf R}={\bf R}(t)-{\bf R}(0)\,$.
Especially (see Introduction), in thermodynamically equilibrium
Boltzmann-Grad gas where {\it a fortiori}\, any ``collective'' or
``hydrodynamic'' contributions to chaotic motion of particles
disappear. At the same time, as far as possible, we want to take in
mind also finite-density and non-equilibrium gas and even liquid.

Instead of ``art of decomposition'' of BP's path into some
constituent parts and ``art of conjecturing'' about their
``probabilities'', we want to follow N.\,Krylov (see Introduction)
and consider a set of mental ``full-scale experiments'' to see how
BP's path as a whole is influenced by artificial controllable
perturbations of both gas and BP. Thus we involve into consideration
statistical correlations between BP's path and gas atoms. At that,
the only ``probability'' to be specified (at our own choosing,
without any conjectures) is initial probabilistic measure in the
space of states of the system. To find other ``probabilities'', one
should analyze a flow of this measure according to the Liouville
equation or equivalent BBGKY equations. But a great deal can be found
from time reversibility of this flow only which is expressed e.g. by
the ``generalized fluctuation-dissipation relations'' (FDR)
\cite{jetp1,jetp2,p}.

Let, firstly, $\,{\bf q}=\{{\bf R},{\bf r}_1,...\,,{\bf r}_N\}$ and
$\,{\bf p}=\{{\bf P},{\bf p}_1,...\,,{\bf p}_N\}\,$ are (canonical)
coordinates and momentums of our system, and $\,H({\bf q},{\bf p})\,$
its Hamiltonian in absence of its perturbations, so that $\,H({\bf
q},-{\bf p})=H({\bf q},{\bf p})\,$. Secondly, the initial
probabilistic measure, $\,\rho_{\,in}({\bf q},{\bf p})\,$, is the
equilibrium canonical one corresponding to this Hamiltonian, that is
$\,\,\rho_{\,in}({\bf q},{\bf p})=\rho_{\,eq}({\bf q},{\bf p})\propto
\exp{[-H({\bf q},{\bf p})/T\,]}\,$ (omitting a normalization factor).
Thirdly, suppose possibility that BP is perturbed by a constant
external force $\,{\bf f}\,$ which is sharply switched on at
$\,t=0\,$, thus at $\,t>0\,$ changing the Hamiltonian to $\,H(t,{\bf
q},{\bf p})= H({\bf q},{\bf p})-{\bf f}\cdot{\bf R}\,$.

Then we can either directly write according to works \cite{jetp1} or
\cite{p} or derive by their methods, like in \cite{p71}, the very
particular but useful enough FDR as follows:
\begin{equation}
\begin{array}{c}
\langle \,A({\bf q}(t),{\bf p}(t))\,B({\bf q}(0),{\bf
p}(0))\,e^{-\,\mathcal{E}(t)/T}\,\rangle_0\,=
\,\,\,\,\,\,\,\,\,\,\,\,\,\,\,\,\,\,\,\,\,\,\,\,\,\,\,\,\,\,\,\,\,\,
\,\,\,\,\,\,\,\,\,\,\,\,\,\,\,\,\,\,\,\,\\\,\,\,\,\,\,\,\,\,\,\,\,
\,\,\,\,\,\,\,\,\,\,\,\,\,\,\,\,\,\,\,\,\,\,\,\,\,\,\,\,\,\,\,\,\,\,\,
\,\,\,\,\,\,\,\,\,\,\,\,\,\,=\,\langle \,B({\bf q}(t),-{\bf
p}(t))\,A({\bf q}(0),-{\bf p}(0))\,\rangle_0\,\label{sim0}
\end{array}
\end{equation}
Here $\,\mathcal{E}(t)={\bf f}\cdot [{\bf R}(t)-{\bf R}(0)]\,$ is
work made by the external force during time interval $\,(0,t)\,$,
$\,A({\bf q},{\bf p})\,$ and $\,B({\bf q},{\bf p})\,$ are ``arbitrary
functions'', the angle brackets $\langle ...\rangle_0\,$ designate
averaging over statistical ensemble of phase trajectories of the
system corresponding to the equilibrium Gibbs ensemble of their
initial conditions, and $\,T\,$ is initial temperature of the system
(or, to be precise, of the ensemble). At any concrete trajectory, of
course, $\,{\bf q}(t)\,$ and $\,{\bf p}(t)\,$ represent solutions to
Hamilton equations corresponding to the Hamiltonian $\,H(t,{\bf
q},{\bf p})= H({\bf q},{\bf p})-{\bf f}\cdot{\bf R}\,$ and hence are
some functions of the force $\,{\bf f}\,$. The equality (\ref{sim0})
holds also if BP has internal degrees of freedom. In absence of
external force, when $\,\mathcal{E}(t)=0\,$, it is so obvious that
even does not need in a proof.

Further it is sufficient to confine ourselves by such particular
choice as
\begin{equation}
\begin{array}{l}
A({\bf q},{\bf p})\,=\,\delta({\bf R}-{\bf R}^{\prime})\,\,\,,\\
B({\bf q},{\bf p})\,=\,\Omega\,\,\delta({\bf R}-{\bf
R}_0)\,\delta({\bf P}-{\bf P}_0)\exp{[\,-\sum_{j\,=1}^N U({\bf
r}_j,{\bf p}_j)/T\,]}\,\\\,\,\,\,\,\,\,\,\,\,\,\,\,\,
\,\,\,\,\,\,\,\,=\,\Omega\,\,\delta({\bf R}-{\bf R}_0)\,\delta({\bf
P}-{\bf P}_0)\prod_{j\,=1}^N [1+\psi({\bf r}_j,{\bf p}_j)\,]
\,\,\,,\label{ch}
\end{array}
\end{equation}
where\, $\,\psi({\bf r},{\bf p})\,\equiv\,\exp{[-\,U({\bf r},{\bf
p})/T\,]}-1\,$.

\subsection{Generating functional of distribution functions}

Under choice (\ref{ch}) in the thermodynamical limit right-hand side
of (\ref{sim0}) takes form
\begin{equation}
\begin{array}{l}
\langle \,B({\bf q}(t),-{\bf p}(t))\,A({\bf q}(0),-{\bf
p}(0))\,\rangle_0\, =\,\mathcal{F}\{t,{\bf R}_0,-{\bf
P}_0,\,\psi\,|{\bf R}^{\prime}\}\,\,\label{rs}
\end{array}
\end{equation}
where
\begin{eqnarray}
\mathcal{F}\{t,{\bf R}_0,{\bf P}_0,\,\psi\,|{\bf
R}^{\prime}\}\,\equiv
\,V_0(t,{\bf R}_0,{\bf P}_0|{\bf R}^{\prime})\,+\nonumber\\
+\sum_{n\,=1}^{\infty } \frac {\nu_{\,0}^n}{n!}\int^n_{r\times p}
F_n(t,{\bf R}_0, {\bf r}_1\,...\,{\bf r}_n,{\bf P}_0,{\bf
p}_1\,...\,{\bf p}_n|{\bf R}^{\prime})\prod_{j\,=1}^n \psi({\bf
r}_j,-{\bf p}_j)\,\label{gf}
\end{eqnarray}
Here symbol $\,\int^n_{r\times p}\,$ denotes integration over $\,n\,$
coordinates $\,{\bf r}_1\,...\,{\bf r}_n\,$ and momentums $\,{\bf
p}_1\,...\,{\bf p}_n\,$\,;\, function\, $\,V_0(t,{\bf R}_0,{\bf
P}_0|{\bf R}^{\prime})\,$ is conditional probability density of
finding BP at $\,t\geq 0\,$ at point $\,{\bf R}_0\,$\, with momentum
$\,{\bf P}_0\,$ under condition that BP had started at $\,t=0\,$ from
point ${\bf R}^{\prime}\,$,\, and $\,F_n(t,{\bf R}_0,{\bf
r}_1\,...\,{\bf r}_n,{\bf P}_0,{\bf p}_1\,...\,{\bf p}_n|{\bf
R}^{\prime})\,$ is joint conditional probability density of this
event and simultaneously finding some atoms at points $\,{\bf
r}_j\,$\, with momentums $\,{\bf p}_j\,$ under the same condition.

In respect to atoms, all the $\,F_n\,$ are complete analogues of
standardly defined non-normalized many-particle DF of infinite gas
\cite{bog}. Instead of normalization, ``asymptotic uncoupling'' of
inter-particle correlations takes place:
\[
\begin{array}{l}
F_n(...\,{\bf r}_k\,...\,{\bf
p}_k\,...\,)\,\rightarrow\,F_{n-1}(...\,{\bf r}_{k-1}, {\bf
r}_{k+1}...\,\,{\bf
p}_{k-1}, {\bf p}_{k+1}...\,)\,G_m({\bf p}_k)\,\,\,,\\
F_1(t,{\bf R}_0,{\bf P}_0,{\bf r}_1,{\bf p}_1 |{\bf
R}^{\prime})\,\rightarrow\, V_0(t,{\bf R}_0,{\bf P}_0|{\bf
R}^{\prime}\,)\,G_m({\bf p}_1)\,\,\,,
\end{array}
\]
when $\,{\bf r}_k\rightarrow \infty\,$ and $\,{\bf r}_1\rightarrow
\infty\,$, respectively, with $\,G_m({\bf p})\,$ being equilibrium
Maxwellian distribution of atomic momentum, $\,G_m({\bf p})=(2\pi
Tm)^{-3/2}\exp{(-{\bf p}^2/2Tm)}\,$, and $\,m\,$ atomic mass. But,
because of initial localization of BP, in respect to BP's variables
all the DF are normalized in literal sense. In particular,
\[
\begin{array}{l}
\int V_0(t,{\bf R}_0,{\bf P}_0|{\bf R}^{\prime})\,d{\bf P}_0
\,=\,V_0(t,{\bf R}_0-{\bf R}^{\prime})\,\,\,,\\
 \int V_0(t,{\bf R}_0-{\bf R}^{\prime})\,d{\bf R}_0\, =\,1\,
\end{array}
\]
In respect to $\,\psi\,$, expression $\,\mathcal{F}\{t,{\bf R}_0,{\bf
P}_0,\,\psi\,|{\bf R}^{\prime}\}\,$ represents generating functional
for these DF quite similar to the functional originally introduced by
Bogolyubov \cite{bog}.

By definition of the average $\,\langle ...\,\rangle_0\,$, all the DF
represent BP in initially thermodynamically equilibrium gas (or, to
be precise, in equilibrium Gibbs ensemble of identical systems ``gas
plus BP''). Correspondingly, initial conditions to them are
\begin{eqnarray}
V_0(0,{\bf R}_0,{\bf P}_0|{\bf R}^{\prime})\,=\,\delta({\bf R}_0-{\bf
R}^{\prime})\,G_M({\bf P}_0) \,\,\,, \label{ic0}
\end{eqnarray}
\begin{equation}
\begin{array}{l}
F_n(0,{\bf R}_0, {\bf r}_1\,...\,{\bf r}_n,{\bf P}_0,{\bf
p}_1\,...\,{\bf p}_n|{\bf R}^{\prime})\,=\, \\
\,\,\,\,\,\,\,\,\,\,\,\,=\,F_n^{(eq)}({\bf r}_1...\,{\bf r}_n|{\bf
R}_0)\,\delta({\bf R}_0-{\bf R}^{\prime})\,G_M({\bf
P}_0)\prod_{j\,=1}^n G_m({\bf p}_j)\,\,\,, \label{ic}
\end{array}
\end{equation}
\[
\begin{array}{c}
\mathcal{F}\{0,\,{\bf R}_0,{\bf P}_0,\,\psi\,|{\bf R}^{\prime}\}\, =
\,\delta({\bf R}_0-{\bf R}^{\prime})\,G_M({\bf P}_0)\,
\mathcal{F}^{(eq)}\{\phi |{\bf R}_0\}\,\,\,,
\end{array}
\]
where
\begin{equation}
\begin{array}{l}
\phi({\bf r})\,=\,\int \psi({\bf r},-{\bf p})\,G_m({\bf p})\,d{\bf
p}\,=\,\int \psi({\bf r},{\bf p})\,G_m({\bf p})\,d{\bf
p}\,\,\,,\label{fi}
\end{array}
\end{equation}
\[
\mathcal{F}^{(eq)}\{\phi |{\bf R}_0\}\,\equiv \,1\,
+\sum_{n\,=1}^{\infty }\frac {\nu_{\,0}^n}{n!}\int^n_r
F_n^{(eq)}({\bf r}_1...\,{\bf r}_n|{\bf R}_0) \prod_{j\,=1}^n
\phi({\bf r}_j)\,\,,
\]
functions $\,F_n^{(eq)}({\bf r}_1...\,{\bf r}_n|{\bf R}_0)\,$ are
conditional equilibrium DF of gas under fixed position of BP,\,
$\,\mathcal{F}^{(eq)}\{\phi |{\bf R}_0\}\,$\, is their generating
functional, and $\,M\,$ means BP's mass.

Of course, all the DF possess translational invariance. Notice also
that, firstly, ratios $\,F_n/V_0\,$ represent conditional DF of gas
under condition that both initial position of BP and its current
state are known. Secondly, al least under condition
\begin{equation}
\begin{array}{l}
\int |\phi({\bf r})|\,d{\bf r}\,<\,\infty\,\, \label{cond}
\end{array}
\end{equation}
the infinite series are converging and hence the functionals are well
defined.

\subsection{BP-gas correlations}

In absence of the external force ($\,{\bf f}=0\,$) just listed DF
describe Brownian motion in gas which all the time remains in exact
thermodynamical equilibrium. Nevertheless, at $\,t>0\,$ not only
$\,F_n\,$ but conditional DF $\,F_n/V_0\,$ too are constantly
changing along with $\,V_0\,$. In principle, their joint evolution is
unambiguously prescribed by the BBGKY equations together with initial
conditions (\ref{ic0}) and (\ref{ic}) \cite{i1,i2,p71,p1}. At that,
differences
\[
\begin{array}{l}
\fl \,F_n(t,{\bf R}_0, {\bf r}_1\,...\,{\bf r}_n,{\bf P}_0,{\bf
p}_1\,...\,{\bf p}_n|{\bf R}^{\prime})\,-\,V_0(t,{\bf R}_0,{\bf
P}_0|{\bf R}^{\prime})\,F_n^{(eq)}({\bf r}_1...\,{\bf r}_n|{\bf
R}_0)\prod_j G_m({\bf p}_j)\,
\end{array}
\]
reflect additional specific statistical correlations between BP and
gas which arise just due to evolution of $\,V_0(t,{\bf R}_0,{\bf
P}_0|{\bf R}^{\prime})\,$. Roughly, an origin of their specificity is
that they are correlations of a current gas state with previous BP's
path accumulated during all the time interval $\,(0,t)\,$. By this
reason we can characterize them as ``historical correlations''.

By tradition, any additions to equilibrium (or quasi-equilibrium) DF
are termed ``correlation functions'' (CF) \cite{re,bog,bal}. We will
apply this term also to functions what describe the ``historical
correlations''. Let us designate CF as $\,V_n(t,{\bf R}_0,{\bf
r}_1\,...\,{\bf r}_n,{\bf P}_0,{\bf p}_1\,...\,{\bf p}_n|{\bf
R}^{\prime})\,$ and introduce them through their generating
functional:
\begin{equation}
\begin{array}{l}
\mathcal{F}\{t,{\bf R}_0,{\bf P}_0,\,\psi\,|{\bf R}^{\prime}\}\,\,=\,
\mathcal{F}^{(eq)}\{\phi |{\bf R}_0\}\,\,\mathcal{V}\{t,{\bf
R}_0,{\bf P}_0,\,\psi\,|{\bf R}^{\prime}\}\,\,\,,
 \label{vdef}
\end{array}
\end{equation}
where $\,\phi=\phi({\bf r})\,$ is expressed through $\,\psi=\psi({\bf
r},{\bf p})\,$ by formula (\ref{fi}), and
\begin{equation}
\begin{array}{l}
\mathcal{V}\{t,{\bf R}_0,{\bf P}_0,\,\psi\,|{\bf R}^{\prime}\}\,
\,=\, \,V_0(t,{\bf R}_0,{\bf P}_0|{\bf R}^{\prime})\,\,+
 \label{vf}
\end{array}
\end{equation}
\[
\,\,\,\,\,\,\,\,\,\,\,+\,\sum_{n\,=1}^{\infty } \frac
{\nu_{\,0}^n}{n!}\int^n_{r\times p} V_n(t,{\bf R}_0, {\bf
r}_1\,...\,{\bf r}_n,{\bf P}_0,{\bf p}_1\,...\,{\bf p}_n|{\bf
R}^{\prime})\prod_{j\,=1}^n \psi({\bf r}_j,-{\bf p}_j)\,
\]
According to this definition, in particular,
\begin{eqnarray}
F_1(t,{\bf R}_0,{\bf r}_1,{\bf P}_0,{\bf p}_1|{\bf
R}^{\prime})\,=\label{cf1} \\
=\, V_0(t,{\bf R}_0,{\bf P}_0|{\bf R}^{\prime})\, F_1^{(eq)}({\bf
r}_1|{\bf R}_0)\,G_m({\bf p}_1)\,+\, V_1(t,{\bf R}_0,{\bf r}_1,{\bf
P}_0,{\bf p}_1|{\bf R}^{\prime})\,\,\,,\nonumber
\end{eqnarray}
where function $\,V_1\,$ represents pair ``historical'' correlation
between BP and atoms.

In view of the asymptotic decoupling of inter-particle correlations,
it is clear that all the CF disappear in initial equilibrium state,
$\,V_n(0\,,...\,)=0\,$ at $\,n> 0\,$, as well as far from BP, i.e.\,
$\,V_n(t\,,...\,{\bf r}_k...\,)\rightarrow 0\,$\, at\, $\,{\bf
r}_k\rightarrow \infty\,$.

\subsection{Main relation between correlation functions and response
of Brownian path to gas perturbations}

Now let us introduce\, $\,\Psi ({\bf q},{\bf
p})=\,\Omega\,\delta({\bf R}-{\bf R}_0)\prod_j [1+\psi({\bf r}_j,{\bf
p}_j)\,]\,$\, and, under the choice (\ref{ch}), consider left side of
(\ref{sim0}) rewriting it as
\begin{equation}
\begin{array}{l}
\fl \langle \,A({\bf q}(t),{\bf p}(t))\,B({\bf q}(0),{\bf
p}(0))\,e^{-\,\mathcal{E}(t)/T}\,\rangle_0\,=
\\ \,\,\,\,\,\,\,
=\,\langle \Psi ({\bf q},{\bf p})\rangle_0\,\, \langle \delta({\bf
R}(t)-{\bf R}^{\prime})\,\delta({\bf P}(0)-{\bf
P}_0)\,\rangle\,\,e^{-\,{\bf f}\cdot [\,{\bf R}^{\,\prime}-\,{\bf
R}_0]/T}\,\,, \label{ls}
\end{array}
\end{equation}
where the brackets $\,\langle ...\,\rangle\,$ are defined by
\[
\begin{array}{c}
\langle \Phi\,\rangle \,\equiv \,\langle \Phi\,\Psi ({\bf q},{\bf
p})\,\rangle_0\,/\,\langle \Psi ({\bf q},{\bf p})\rangle_0
\end{array}
\]
with $\,\Phi\,$ being arbitrary functional of the system's phase
trajectory.

Evidently, $\,\langle ...\,\rangle\,$ in fact represents averaging
over new statistical ensemble of initial conditions, with new
probabilistic measure
\[
\begin{array}{c}
\fl \rho_{\,in}({\bf q},{\bf p})\,\propto\,\, \rho_{\,eq}({\bf
q},{\bf p})\Psi ({\bf q},{\bf p})\,\propto\, \delta({\bf R}-{\bf
R}_0)\,\exp{[-H({\bf q},{\bf p})/T-\sum_j U({\bf q}_j,{\bf
p}_j)/T\,]}\,
\end{array}
\]
(again normalizing coefficients are omitted). Formally this measure
can be viewed as also a canonical equilibrium one but in presence of
generalized (momenta-dependent) external potential $\,U({\bf q},{\bf
p})=-T\,\ln{[1+\psi({\bf q},{\bf p})\,]}\,$. In fact, of course, this
is thermodynamically non-equilibrium measure since system's
Hamiltonian does not include such potential. Hence, second right-hand
average in (\ref{ls}) is nothing but
\begin{equation}
\begin{array}{l}
\fl \,\,\,\,\,\,\,\,\,\,\,\,\,\,\,\,\, \langle \delta({\bf R}(t)-{\bf
R}^{\prime})\,\delta({\bf P}(0)-{\bf P}_0)\,\rangle\,=\,V\{t,{\bf
R}^{\,\prime}|\psi ,{\bf R}_0 ,{\bf P}_0\}\,G_M({\bf
P}_0)\,\,\,,\label{ls1}
\end{array}
\end{equation}
where $\,V\{t,{\bf R}^{\,\prime}|\psi ,{\bf R}_0 ,{\bf P}_0\}\,$ is
conditional probability density of finding BP at point $\,{\bf
R}^{\,\prime}\,$ under conditions that initially, at $\,t=0\,$, it
was located at point $\,{\bf R}_0\,$ with momentum $\,{\bf P}_0\,$
while gas was in such non-equilibrium spatially-nonuniform state what
would be equilibrium under external potential $\,U({\bf q},{\bf
p})\,$. Noticing, besides, that
\[
\begin{array}{l}
\langle \Psi ({\bf q},{\bf p})\rangle_0\,=\, \mathcal{F}^{(eq)}\{\phi
|{\bf R}_0\}\,\,,
\end{array}
\]
(again with $\,\phi\,$ expressed through $\,\psi\,$ by (\ref{fi})),
we can write
\begin{equation}
\begin{array}{l}
\fl \langle \,A({\bf q}(t),{\bf p}(t))\,B({\bf q}(0),{\bf
p}(0))\,e^{-\,\mathcal{E}(t)/T}\,\rangle_0\,=
\\ \,\,\,\,\,
=\,V\{t,{\bf R}^{\,\prime}|\psi ,{\bf R}_0 ,{\bf P}_0\}\,\,G_M({\bf
P}_0)\,\,e^{-\,{\bf f}\cdot [\,{\bf R}^{\,\prime}-\,{\bf
R}_0]/T}\,\,\mathcal{F}^{(eq)}\{\phi |{\bf R}_0\}\, \label{ls2}
\end{array}
\end{equation}

Equivalently, instead of the artificial external potential $\,U({\bf
q},{\bf p})\,$, the non-equilibrium ensemble which has arisen can be
characterized by corresponding conditional mean densities of atoms in
the $\,\mu$-space and coordinate space,
\[
\fl \,\,\,\,\,\,\,\,\mu\{{\bf r},{\bf p}|\,\psi ,{\bf R}_0
\}\,\equiv\,\langle \,\sum_j\delta({\bf r}_j-{\bf r})\delta({\bf
p}_j-{\bf p})\,\rangle\, =\,\nu\{{\bf r}|\,\phi ,{\bf
R}_0\}\,\,G_m({\bf p})\,\frac {1+\psi({\bf r},{\bf p})}{1+\phi({\bf
r})}\, \,,
\]
\begin{equation}
\fl \,\,\,\,\,\,\,\,\,\,\,\,\,\,\,\,\,\,\,\,\,\nu\{{\bf r}|\,\phi
,{\bf R}_0 \}\,\equiv\,\langle \,\sum_j\delta({\bf r}_j-{\bf
r})\,\rangle\, =\, [1+\phi({\bf r})]\,\,\frac {\delta
\ln\mathcal{F}^{(eq)}\{\phi |{\bf R}_0 \}}{\delta\phi({\bf r})}
\label{n}
\end{equation}
At that, $\,\mu\{{\bf r},{\bf p}|\,\psi ,{\bf R}_0 \}/\nu_0\,$ has
the sense of initial one-particle DF of atoms.

At last, combining formulas (\ref{sim0}), (\ref{rs}), (\ref{vdef}),
(\ref{vf}) and (\ref{ls2}), we come to formally exact relation
\begin{equation}
V\{t,{\bf R}^{\,\prime}|\psi ,{\bf R}_0 ,{\bf P}_0\}\,\,G_M({\bf
P}_0)\,\,\,e^{-\,{\bf f} \cdot[\,{\bf R}^{\,\prime}-\,{\bf R}_0]/T}
\,=\label{r}
\end{equation}
\[
\fl =V_0(t,{\bf R}_0,-{\bf P}_0|{\bf R}^{\prime})
+\sum_{n\,=1}^{\infty } \frac {\nu_{\,0}^n}{n!}\int^n_{r\times p}
V_n(t,{\bf R}_0,{\bf r}_1...{\bf r}_n,-{\bf P}_0,{\bf p}_1\,...{\bf
p}_n|{\bf R}^{\,\prime}) \prod_{j\,=1}^n \psi({\bf r}_j,-{\bf p}_j)\,
\]
It connects, from one (left) hand, probability distribution of BP's
path in initially non-equilibrium nonuniform gas and, from the other
(right) hand, probability distribution of BP's path, along with
generating functional of correlations between this previously
accumulated path and current BP's environment, in initially
equilibrium uniform gas. In case $\,{\bf f}=0\,$, therefore,
right-hand side of (\ref{r}) represents wholly equilibrium Brownian
motion.

In other words, relation (\ref{r}) connects two sorts of ``full-scale
experiments'': one on susceptibility of Brownian motion to
perturbations of medium where it takes place, and another on its
correlations with thermal fluctuations in the medium. In such sense,
(\ref{r}) is typical generalized FDR or ``generalized Onsager
relation''.

At $\,\psi({\bf r},{\bf p})=0\,$, clearly, (\ref{r}) turns into
\begin{equation}
\begin{array}{l}
\fl \,\,\,\,\,\,\,\,\,\,\,\,\,\,\,\,\,\,\,\,V_0(t,{\bf
R}^{\,\prime}|{\bf R}_0 ,{\bf P}_0)\,\,G_M({\bf P}_0)\,\,\,e^{-\,{\bf
f} \cdot[\,{\bf R}^{\,\prime}-\,{\bf R}_0]/T} \,=\,V_0(t,{\bf
R}_0,-{\bf P}_0|{\bf R}^{\prime})\,\,\,,\label{r00}
\end{array}
\end{equation}
where $\,V_0(t,{\bf R}^{\,\prime}|{\bf R}_0 ,{\bf P}_0\}\,$ is
density of probability to find BP (in the same gas as on the left) at
point $\,{\bf R}^{\,\prime}\,$ under condition that it started from
point $\,{\bf R}_0\,$ with momentum $\,{\bf P}_0\,$. Integration over
this momentum yields the classical FDR \cite{jetp2,p,bk2,bk3}
\begin{equation}
\,\,\,V_0(t,\Delta {\bf R})\,\,e^{-\,{\bf f} \cdot \Delta {\bf R}/T}
\,=\,V_0(t,-\Delta {\bf R})\,\,\,, \label{r0}
\end{equation}
where $\,\Delta {\bf R}\equiv {\bf R}^{\,\prime}-{\bf R}_0\,$\, and
\[
\,\,\,V_0(t,{\bf R}_0-{\bf R}^{\,\prime})\,=\,\int V_0(t,{\bf
R}_0,{\bf P}_0|{\bf R}^{\prime})\,d{\bf P}_0\,
\]

Of course, all the DF and CF are dependent on the mean gas density
$\,\nu_{\,0}\,$ (and on the force $\,{\bf f}\,$ if any) but for
brevity in (\ref{r}), as well as before it and almost everywhere
below, corresponding arguments are omitted.

\subsection{Quasi-uniform gas perturbations,
spatial correlations and virial expansion}

Let, firstly, the gas perturbation does not change velocity
distribution of atoms, that is represents pure density perturbation,
$\,\psi({\bf r},{\bf p})=\phi({\bf r})\,$. Secondly, $\,\phi({\bf
r})=\phi =\,$const\, inside some sphere $\,|{\bf r}-{\bf R}_0|<
\xi\,$ and vanishes outside it in some suitable way (it should be
underlined that nothing impedes choosing perturbations to be
correlated with the point $\,{\bf R}_0\,$). Since, according to
(\ref{cond}), $\,\phi({\bf r})\,$ is absolutely integrable, radius
$\,\xi\,$ must be finite. But it can be as large as we want. At that,
factual perturbation of gas equilibrium initially is located at
$\,|{\bf r}-{\bf R}_0|> \xi\,$.

For example, we may take $\,\xi = k\,v_s t_0\,$, where $\,v_s\,$ is
speed of sound in our gas (or liquid), $\,t_0\,$ is maximal duration
of our ``full-scale experiments'', and $\,k>2\,$. Then, if the
external force is not too strong (so that velocity of BP's drift
induced by the external force is small as compared with $\,v_s\,$),
we can be sure that at $\,t<t_0\,$ the Brownian motion mentioned in
left part of (\ref{r}) practically takes place in equilibrium uniform
gas with a constant mean density $\,\nu =\,$const\, what corresponds
to $\,\phi =\,$const\,. Indeed, under mentioned conditions $\,|{\bf
R}(t)-{\bf R}(0)|<v_s t\,$ while radial approach of inner front of
the gas perturbation (let even Mach front) to the point $\,{\bf
R}(0)={\bf R}_0\,$ is not greater than $\,v_s t\,$. Hence,
undoubtedly BP does not feel the front and moves as it was in
spatially uniform (and thus equilibrium) media at least till
$\,t<t_0\,$. This can be named ``quasi-uniform perturbation'' of gas.

Now, at $\,t<t_0\,$, in fact both parts of (\ref{r}) describe
Brownian motions in (initially) uniform equilibrium gases but with
different values of density. On the right it is $\,\nu_0\,$ while on
the left it equals to $\,\nu\{{\bf r}|\phi ,{\bf R}_0 \}\,$ taken far
from BP, at $\,|{\bf r}-{\bf R}_0|\gg r_B\,$, with $\,r_B\,$ being
radius of pair interaction between BP and atoms. We will denote this
value simply as $\,\nu\,$. According to (\ref{n}), it is definite
function of $\,\phi \,$ and the seed density $\,\nu_0\,$:
\begin{equation}
\begin{array}{l}
\nu =\nu(\nu_{\,0}\,,\phi )\,=\,\nu_{\,0}\,(1+\phi
)\,\{\,1+\nu_{\,0}\phi \int [\,F^{(eq)}_2({\bf r})-1]\,+\,\,\,\,
\label{n1}
\end{array}
\end{equation}
\[
\fl \,\,\,\,+\,\frac {\nu_{\,0}^2\phi^2}{2} \int_r^{\,2}
[\,F^{(eq)}_3({\bf r}_1,{\bf r}_2,0)-F^{(eq)}_2({\bf
r}_1)-F^{(eq)}_2({\bf r}_2)-F^{(eq)}_2({\bf r}_1-{\bf r}_2)+2\,]\,
+...\,\}
\]
Here $\,F^{(eq)}_2({\bf r})\,$ and $\,F^{(eq)}_3({\bf r}_1,{\bf
r}_2,{\bf r}_3)\,$ are standard pair and triple DF of equilibrium gas
with density $\,\nu_{\,0}\,$ (i.e. functions what follow from
$\,F^{(eq)}_n({\bf r}_1...{\bf r}_n|{\bf R}_0)\,$ at $\,{\bf
R}_0\rightarrow \infty\,$).

Correspondingly, the functional $\,V\{t,{\bf R}^{\prime}|\phi ,{\bf
R}_0,{\bf P}_0  \}\,$ simplifies to mere function, and, integrating
left side of (\ref{r}) over $\,{\bf P}_0\,$, for $\,t<t_0\,$ we can
write
\[
\begin{array}{l}
\int V\{t,{\bf R}^{\prime}|\phi ,{\bf R}_0,{\bf P}_0 \}\,G_M({\bf
P}_0)\,d{\bf P}_0\,=\,V_0(t,{\bf R}^{\prime}-{\bf R}_0\,;\,\nu
)\,\,\,,
\end{array}
\]
where $\,V_0\,$ has exactly the same sense as $\,V_0\,$ on the
right-hand side of (\ref{r}) (after its integration over $\,{\bf
P}_0\,$), and we introduced the density argument $\,\nu \,$, so that
\[
\begin{array}{l}
V_0(t,{\bf R}_0-{\bf R}^{\prime}\,;\,\nu =\nu_0\,)\,=\,V_0(t,{\bf
R}_0-{\bf R}^{\prime})
\end{array}
\]

On right-hand side of (\ref{r}), under the formulated conditions,
{\it a fortiori} none correlations between BP and gas might propagate
out of the sphere $\,|{\bf r}-{\bf R}_0|< \xi\,$. Therefore in all
the integrals $\,\phi({\bf r})\,$ can be replaced by the constant.
Consequently, relation (\ref{r}), after its integration over $\,{\bf
P}_0\,$, transforms into
\begin{equation}
\fl \,\,\,V_0(t,\Delta{\bf R}\,;\,\nu(\nu_{\,0},\psi
))\,\,e^{\,-\,{\bf f}\cdot\Delta {\bf R}/T}\,=\,V_0(t,-\Delta{\bf
R}\,;\,\nu_{\,0}) \,+\, \sum_{n\,=\,1}^{\infty } \,\frac {\phi^n}{n!}
\,V_n(t,-\Delta{\bf R}\,;\,\nu_{\,0})\,\label{r2}
\end{equation}
functions $\,V_n\,$ at $\,n>0\,$ defined by
\begin{equation}
\fl \,\,\,\,\,\,\,\,\,\,\,\,V_n(t,{\bf R}_0-{\bf
R}^{\prime}\,;\,\nu_{\,0})\,=\,\nu_{\,0}^n\,\int^n_{r\times p} \int
V_n(t,{\bf R}_0,{\bf r}_1...\, {\bf r}_n,{\bf P}_0,{\bf p}_1...\,
{\bf p}_n\,|\,{\bf R}^{\prime})\,d{\bf P}_0 \label{clim0}
\end{equation}
Combining (\ref{r2}) with FDR (\ref{r0}) (which, of course, is valid
at any density if equal on both sides), we obtain
\begin{equation}
\fl \,\,\,\,\,\,\,\,\,\,\,\,\,\,\,\,\,\,\,\,\, V_0(t,\Delta{\bf
R}\,;\,\nu(\nu_{\,0},\psi ))\,=\,V_0(t,\Delta{\bf R}\,;\,\nu_{\,0})
\,+\, \sum_{n\,=\,1}^{\infty } \,\frac {\phi^n}{n!}
\,V_n(t,\Delta{\bf R}\,;\,\nu_{\,0})\,\,\,,\label{ve}
\end{equation}
where now $\,\Delta{\bf R}\equiv {\bf R}_0-{\bf R}^{\prime}\,$.

Relation (\ref{ve}) thoroughly may be qualified as ``virial
expansion'' of the BP's path probability distribution,
$\,V_0(t,\Delta{\bf R}\,;\,\nu )\,$, similarly to well-known virial
expansions of thermodynamical quantities \cite{ll1} or kinetic
coefficients \cite{ll2}. But there is significant mathematical
difference between them: the two latter expand over absolute value of
density $\,\nu\,$ while the former in fact over its relative value
$\,\nu/\nu_0\,$.

\subsection{First-order relations}

First-order, in respect to $\,\phi\,$, terms of (\ref{ve}) produce
\begin{eqnarray}
\fl \,\,\widetilde{\nu }_0\,\frac {\partial V_0(t,\Delta {\bf
R}\,;\,\nu_{\,0})}{\partial \nu_{\,0}}\, = \,V_1(t,\Delta {\bf
R}\,;\,\nu_{\,0})\,= \,\nu_{0}\int_{r\times p}\int V_1(t,{\bf
R}_0,{\bf P}_0,{\bf r}_1,{\bf p}_1|{\bf R}^{\prime})\,d{\bf P}_0 \,
\label{r21}
\end{eqnarray}
Here $\,\Delta {\bf R}={\bf R}_0-{\bf R}^{\prime}\,$ and, according
to (\ref{n}) and/or (\ref{n1}),
\begin{eqnarray}
\fl \,\,\,\,\,\,\,\,\,\,\,\,\widetilde{\nu }_0\,\equiv \,\left [\frac
{\partial \nu (\nu_{\,0},\psi )}{\partial \psi }\right ]_{\psi =0}
=\,\nu_0 +\nu_0^2\,\int [F_2^{(eq)}({\bf r})-1\,]\,d{\bf
r}\,=\,\nu_{\,0}\,T\left(\frac {\partial \nu_{\,0}}{\partial
P}\right)_T\, \label{n2}
\end{eqnarray}
Here, of course, $\,F_2^{(eq)}({\bf r})\,$ is some function of
$\,\nu_0\,$. The last equality in (\ref{n2}), with $\,P\,$ standing
for the pressure and the bracket representing compressibility, is
well known from statistical thermodynamics \cite{ll1}.

Notice that the function $\,V_1(t,\Delta {\bf R};\nu_{\,0})\,$ can be
interpreted as total pair correlation of BP with whole gas but
counted per (elementary volume $\,\nu_0^{-1} \,$ what falls at) one
gas atom. Similar interpretation is applicable to higher-order
integral correlations (\ref{clim0}).

Returning to our basic relation (\ref{r}) in the full phase space, in
the first order with respect to $\,\psi\,$ we have
\begin{equation}
\fl  \,\,\,\,\,\,\,\,\,\,\,\,\,\left[\,\frac {\delta V\{t,{\bf
R}^{\,\prime}|\psi ,{\bf R}_0 ,{\bf P}_0\}}{\delta \psi({\bf r},{\bf
p})}\,\right ]_{\psi \,=0}\,G_M({\bf P}_0)\,\,\,e^{-\,{\bf f}
\cdot[\,{\bf R}^{\,\prime}-\,{\bf R}_0]/T} \,=\, \label{fo}
\end{equation}
\[
\begin{array}{c}
\,\,\,\,\,\,\,\,\,\,\,\,\,\,\,\,\,\,\,\,\,\,\,\,\,\,\,
\,\,\,\,\,\,\,\,\,\,\,\,\,\,\,\,\,\,\,\,\,\,\,\,\,\,
\,\,\,\,\,\,\,\,\,\,\,\, =\,\nu_{\,0}\,V_1(t,{\bf R}_0,{\bf r},-{\bf
P}_0,-{\bf p}|{\bf R}^{\,\prime})
\end{array}
\]
With the help of (\ref{n}), variational derivative with respect to
$\,\psi\,$ can be replaced by that with respect to density of atoms
in $\,\mu$-space:
\begin{equation}
\fl \left[\,\frac {\delta V}{\delta \psi({\bf r},{\bf p})}\,\right
]_{\psi \,=0}\,=\,\nu_{\,0}G_m({\bf p})\,F_1^{(eq)}({\bf r}|{\bf
R}_0)\,\frac {\delta V}{\delta \mu({\bf r},{\bf p})}\,+\,\nu_{\,0}^2
G_m({\bf p})\,\times \label{fo1}
\end{equation}
\[
\fl \,\,\,\,\,\,\,\,\,\,\times\,\int\int G_m({\bf
p}^{\,\prime})\,[\,F_2^{(eq)}({\bf r},{\bf r}^{\,\prime}|{\bf
R}_0)-F_1^{(eq)}({\bf r}|{\bf R}_0)\,F_1^{(eq)}({\bf
r}^{\,\prime}|{\bf R}_0)\,] \,\,\frac {\delta V}{\delta \mu({\bf
r}^{\,\prime},{\bf p}^{\,\prime})}\,\,d{\bf r}^{\,\prime} d{\bf
p}^{\,\prime}
\]
with the right-hand variational derivatives taken at $\,\mu({\bf
r},{\bf p})= \nu_{\,0}G_m({\bf p})F_1^{(eq)}({\bf r}|{\bf R}_0)\,$,
i.e. at equilibrium gas with the seed density $\,\nu_{\,0}\,$. This
generalizes (\ref{n2}) to arbitrary non-uniform perturbations.

The two latter formulas establish quite rigorous connection of, from
one hand, pair correlation between previous path of BP and current
state of the medium where it is walking, and, from the other hand,
linear susceptibility of the BP's path probability distribution to
weak perturbations of the medium.

\section{Ranges of spatial statistical correlations between medium
and Brownian particle and restrictions on its path probability
distribution}

Our previous results clearly show that, firstly, historical
correlations between BP and medium (gas or liquid) certainly exist,
i.e. are not zeros. Secondly, their total values (integrated over all
momenta and relative distances between atoms and BP as in
(\ref{clim0})) quite definitely reflect sensitivity of the
probabilistic law of Brownian motion to change of state of the
medium, first of all, to density of its atoms.

Indeed, let the medium represents three-dimensional weakly non-ideal
(``dilute'') gas in equilibrium ($\,{\bf f}=0\,$). Then, undoubtedly,
an increase of gas density must lead to constriction of the
distribution $\,V_0(t,\Delta {\bf R};\nu)\,$. Thus its density
derivative\, $\,\partial V_0(t,\Delta {\bf R}\,;\,\nu_{\,0})/\partial
\nu_{\,0}\,$\, is positive at sufficiently small $\,|\Delta {\bf
R}|\,$ but negative at relatively large $\,|\Delta {\bf R}|\,$.
Since, undoubtedly again, auto-correlation of BP's velocity is
integrable and therefore BP's chaotic walk is organized as
``diffusion'' considered already in \cite{ae1,ae2}, i.e.\, $\,\Delta
{\bf R}^2(t)\propto Dt\,$\,,\, $\,\int \Delta {\bf R}^2\,V_0(t,\Delta
{\bf R};\nu_{\,0})\,d\Delta {\bf R}\,=6Dt\,$\,, a bound what
separates ``small'' and ``large'' values of $\,|\Delta {\bf R}|\,$ is
of order of $\,2\sqrt{Dt}\,$. Hence, we can state that $\,\partial
V_0(t,\Delta {\bf R}\,;\,\nu_{\,0})/\partial \nu_{\,0}\,<0\,$\, when
$\,z\equiv \Delta {\bf R}^2/4Dt\,$ is significantly greater then
unit.

This implies, according to (\ref{r21}), that total pair BP-gas
correlation, $\,V_1(t,\Delta {\bf R}\,;\,\nu_{\,0})\,$, also is
negative at large $\,z\,$, and thus $\,V_1(t,{\bf R}_0,{\bf P}_0,{\bf
r}_1,{\bf p}_1|{\bf R}^{\prime})\,$ somewhere is negative. But it can
not be ``too negative'', because its contribution to right side of
(\ref{cf1}) should not make its left side (which represents
probability distribution) negative. This means, obviously, that the
integral $\,V_1(t,\Delta {\bf R}\,;\,\nu_{\,0})\,$ and thus the
derivative $\,\nu_{\,0}\,\partial V_0(t,\Delta {\bf
R}\,;\,\nu_{\,0})/\partial \nu_{\,0}\,$ is bounded below by some
negative whose absolute value is proportional to $\,V_0(t,\Delta {\bf
R};\nu_{\,0})\,$. Such restriction of the derivative, in its turn,
means that tails of $\,V_0(t,\Delta {\bf R};\nu_{\,0})\,$ at $\,z\gg
1\,$ can not decrease in too fast way. Anyhow, possibility of
exponential decrease of $\,V_0(t,\Delta {\bf R};\nu_{\,0})\,$ at
$\,z\gg 1\,$ and hence the Gaussian asymptotic (\ref{vg}) become
under question. Let us consider this suspicion carefully.

\subsection{Restriction on the BP's path distribution tails. Estimate 1}

At any $\,t\,$, $\,{\bf R}^{\,\prime}\,$, $\,{\bf R}_0\,$, $\,{\bf
P}_0\,$ and $\,{\bf p}\,$\,, let
\[
h(t,{\bf R}_0-{\bf R}^{\,\prime},{\bf P}_0,{\bf p})\,=\,\min_{{\bf
r}}\,{V_1(t,{\bf R}_0,{\bf r},{\bf P}_0,{\bf p}|{\bf R}^{\,\prime})}
\]
Assume that at given $\,t\,$, $\,{\bf R}^{\,\prime}\,$, $\,{\bf
R}_0\,$, $\,{\bf P}_0\,$ and $\,{\bf p}\,$\, integral of
$\,V_1(t,{\bf R}_0,{\bf r},{\bf P}_0,{\bf p}|{\bf R}^{\,\prime})\,$
over $\,{\bf r}\,$ is negative (i.e. $\,V_1(t,\Delta {\bf
R}\,;\,\nu_{\,0})<0\,$). This means that $\,h(t,{\bf R}_0-{\bf
R}^{\,\prime},{\bf P}_0,{\bf p})\,$\, also is negative, and therefore
we can introduce effective volume occupied by negative pair
correlation, $\,\Omega_{neg}(t,\Delta {\bf R},{\bf P}_0,{\bf p})\,$,
by means of
\begin{equation}
\Omega_{neg}(t,\Delta {\bf R},{\bf P}_0,{\bf p})\,\,\equiv\,\frac
{\int V_1(t,{\bf R}_0,{\bf P}_0,{\bf r},{\bf p}|{\bf
R}^{\prime})\,d{\bf r} }{h(t,\Delta {\bf R},{\bf P}_0,{\bf p})}
\,\label{ev}
\end{equation}
Clearly, this is lower boundary of volumes what could be reasonably
attributed to the negative correlation. If, in opposite, the integral
over $\,{\bf r}\,$ is positive, in the same sense it is reasonable to
assign $\,\Omega_{neg}(t,\Delta {\bf R},{\bf P}_0,{\bf p})=0\,$. Then
in any case we can write
\begin{equation}
\begin{array}{c}
\fl \,\,\,\,\,\,\,\,\,\,\,\,\,\,\,\,\,\,\,\,\,\,\,\,\,\,\,\,\,\, \int
V_1(t,{\bf R}_0,{\bf P}_0,{\bf r},{\bf p}|{\bf R}^{\prime})\,d{\bf
r}\,\geq\,\Omega_{neg}(t,\Delta {\bf R},{\bf P}_0,{\bf
p})\,h(t,\Delta {\bf R},{\bf P}_0,{\bf p}) \label{in1}
\end{array}
\end{equation}

Next, pay our attention to identity (\ref{cf1}). Since the pair DF
$\,F_1\,$ represents a probability density, it must be non-negative,
$\,F_1(t,{\bf R}_0,{\bf r},{\bf P}_0,{\bf p}|{\bf R}^{\prime})\geq
0\,$. Hence,
\begin{equation}
\begin{array}{c}
h(t,\Delta {\bf R},{\bf P}_0,{\bf p})\,\,\geq\,\,-\,V_0(t,{\bf
R}_0,{\bf P}_0|{\bf R}^{\prime})\,\,G_m({\bf p})\,
\max{F_1^{(eq)}}\,\, \label{in2}
\end{array}
\end{equation}
with $\,\max\,{F_1^{(eq)}}\,$ denoting maximum of all values of
$\,F_1^{(eq)}({\bf r}|{\bf R}_0)\,$ at those points $\,{\bf r}\,$
where $\,V_1(t,{\bf R}_0,{\bf r},{\bf P}_0,{\bf p}|{\bf
R}^{\,\prime})\,$ takes its minimum value $\,h(t,{\bf R}_0-{\bf
R}^{\,\prime},{\bf P}_0,{\bf p})\,$. This inequality will be even
more so valid if replace $\,\max\,{F_1^{(eq)}}\,$ by absolute maximum
of $\,F_1\,$:\, $\,\max{F_1^{(eq)}}\rightarrow \max_{\,{\bf
r}}{F_1^{(eq)}({\bf r}|{\bf R}_0)}\,$.

It is useful to notice that $\,F_1(t,{\bf R}_0,{\bf r}={\bf R}_0,{\bf
P}_0,{\bf p}|{\bf R}^{\prime})= 0\,$,\, $\,F_1^{(eq)}({\bf R}_0|{\bf
R}_0)=0\,$\, and therefore $\,V_1(t,{\bf R}_0,{\bf r}={\bf R}_0,{\bf
P}_0,{\bf p}|{\bf R}^{\prime})= 0\,$,\, if BP and an atom can not be
located at same point. Consequently, always $\,\,h(t,\Delta {\bf
R},{\bf P}_0,{\bf p})\,\leq\,0\,$.

Combining inequalities (\ref{in1}) and (\ref{in2}) and relation
(\ref{r21}), we come to inequality
\begin{eqnarray}
\fl \,\,\,\,\,\,\,\,\,\,\,\,\widetilde{\nu }_0\,\frac {\partial
V_0(t,\Delta {\bf R}\,;\,\nu_{\,0})}{\partial \nu_{\,0}}\, \geq
\,-\,\nu_{0}\,
\max{F_1^{(eq)}}\,\times  \label{in3} \\
\,\,\,\,\,\,\,\,\,\,\,\,\,\,\,\,\,\,\,\,\,\,\,\,\times\,\int \int
\Omega_{neg}(t,\Delta {\bf R},{\bf P}_0,{\bf p})\,V_0(t,{\bf
R}_0,{\bf P}_0|{\bf R}^{\prime})\,\,G_m({\bf p})\,d{\bf p}\,d{\bf
P}_0 \, \nonumber
\end{eqnarray}
It remains to uncouple the joint distribution of BP's path and
momentum:
\[
\begin{array}{c}
V_0(t,{\bf R}_0,{\bf P}_0|{\bf R}^{\prime})\,=\,\,V_0(t,\Delta {\bf
R}\,;\,\nu_{\,0})\,\,G(t,{\bf P}_0|\Delta {\bf R})\,\,\,,
\end{array}
\]
where last multiplier is conditional probability distribution of the
momentum under given path value, and introduce conditional average
correlation volume
\begin{eqnarray}
\fl \,\,\,\,\,\,\,\,\,\,\,\,\,\,\,\,\overline{\Omega}_{neg}(t,\Delta
{\bf R})\,=\, \int \int \Omega_{neg}(t,\Delta {\bf R},{\bf P}_0,{\bf
p})\,\,G(t,{\bf P}_0|\Delta {\bf R})\,G_m({\bf p})\,d{\bf p}\,d{\bf
P}_0 \, \label{av}
\end{eqnarray}
After that inequality (\ref{in3}) takes the form
\begin{eqnarray}
\fl \,\,\,\,\,\,\,\,\,\,\,\,\,\,\,\,\widetilde{\nu }_0\,\frac
{\partial V_0(t,\Delta {\bf R}\,;\,\nu_{\,0})}{\partial
\nu_{\,0}}\,\, +\,\nu_{\,0}\,
\max{F_1^{(eq)}}\,\,\overline{\Omega}_{neg}(t,\Delta {\bf
R})\,V_0(t,\Delta {\bf R}\,;\,\nu_{\,0})\,\geq\,0 \label{in4}
\end{eqnarray}
of declared restriction on rate of change of $\,V_0(t,\Delta {\bf
R})\,$. Before discussing it, consider another variant of the
restriction \cite{lp}.

\subsection{Restriction on the BP's path distribution tails. Estimate 2}

Let $\,\Omega \,$ denotes at once a finite region in the $\,{\bf
r}$-space and volume of this region, presuming that it is centered
near point $\,{\bf R}_0\,$. Introduce $\,\Omega
(\delta)=\Omega(t,\Delta {\bf R},{\bf P}_0,{\bf p},\delta)\,$ as
minimum of all those regions $\,\Omega \,$ what satisfy
\begin{eqnarray}
\left |\,\int_{\Omega} V_1(t,{\bf R}_0,{\bf r},{\bf P}_0,{\bf p}|{\bf
R}^{\prime}\,)\,d{\bf r}\,-\,\int V_1(t,{\bf R}_0,{\bf r},{\bf
P}_0,{\bf p}|{\bf R}^{\prime}\,)\,d{\bf r}\,\right |\,<\,\nonumber \\
\,\,\,\,\,\,\,\,\,\,\,\,\,\,\,\,\,\,<\,\,\delta\,\,\left |\,\int
V_1(t,{\bf R}_0,{\bf r},{\bf P}_0,{\bf p}|{\bf R}^{\prime}\,)\,d{\bf
r}\,\right |\,\nonumber
\end{eqnarray}
with some fixed $\,0< \delta < 1\,$. That is
$\,\Omega_{cor}(\delta)\,$ represents minimal region containing at
least $\,100\,(1-\delta )\,$ percents of the total pair correlation.

From the other hand, integrate identity (\ref{cf1}) over $\,{\bf r}_1
\in \Omega (\delta)\,$. In view of the non-negativeness of
$\,F_1(t,{\bf R}_0,{\bf r}_1,{\bf P}_0,{\bf p}_1|{\bf R}^{\prime})\,$
result must be non-negative:
\begin{eqnarray}
\fl \,\,\,\int_{\,\Omega(\delta)} F_1^{(eq)}({\bf r}|{\bf
R}_0)\,d{\bf r}\,\,\,V_0(t,{\bf R}_0,{\bf P}_0|{\bf
R}^{\prime})\,G_m({\bf p})\, +\,\int_{\,\Omega(\delta)} V_1(t,{\bf
R}_0,{\bf r},{\bf P}_0,{\bf p}|{\bf R}^{\prime}) \,d{\bf
r}\,\geq\,0\,\,\nonumber
\end{eqnarray}
It is easy to verify that these two inequalities together imply a
more interesting one:
\begin{eqnarray}
\fl \,\,\,\,\,\,\,\,\,\,\,\,\,\,\,\,\,\,\,\,\,\,\,
\,\int_{\,\Omega(\delta)} F_1^{(eq)}({\bf r}|{\bf R}_0)\,d{\bf
r}\,\,\,V_0(t,{\bf R}_0,{\bf P}_0|{\bf R}^{\prime})\,G_m({\bf p})\, +
\label{in} \\
\,\,\,\,\,\,\,\,\,\,\,\,\,\,\,\,\,\,\,\,\,\,\,\,\,\,\,\,\,
\,\,\,\,\,\,\,\,\,\,\,\,\,\,\,\,+\,\, (1-\delta\,)\int V_1(t,{\bf
R}_0,{\bf r},{\bf P}_0,{\bf p}|{\bf R}^{\prime}) \,d{\bf
r}\,\,\geq\,\,0\,\,\nonumber
\end{eqnarray}
Next, again let us replace here $\,F_1^{(eq)}\,$ by
$\,\max\,{F_1^{(eq)}}\,$ (which even improves the inequality), then
perform integration over momentums, apply relation (\ref{r21}) and
divide all by $\,(\,1-\delta )\,$. This yields
\begin{eqnarray}
\fl \,\,\,\,\,\,\,\,\,\,\,
\nu_{0}\,\max\,{F_1^{(eq)}}\,\,\overline{\Omega}(t,\Delta {\bf
R},\,\delta)\,\,V_0(t,\Delta {\bf
R}\,;\,\nu_{\,0})\,+\,\,\widetilde{\nu }_0\,\frac {\partial
V_0(t,\Delta {\bf R}\,;\,\nu_{\,0})}{\partial \nu_{\,0}}\,\, \geq\,0
\label{in5}
\end{eqnarray}
with characteristic average pair correlation volume defined by
\begin{eqnarray}
\fl \,\,\,\overline{\Omega}(t,\Delta {\bf R},\,\delta)\,=\,
(\,1-\delta )^{-\,1}\,\int \int \Omega(t,\Delta {\bf R},{\bf
P}_0,{\bf p},\delta)\,\,G(t,{\bf P}_0|\Delta {\bf R})\,G_m({\bf
p})\,d{\bf p}\,d{\bf P}_0 \, \label{av1}
\end{eqnarray}

Unlike (\ref{av}) this characteristic volume has free parameter
$\,0<\delta <1\,$. At $\,\delta \ll 1\,$ almost all pair correlation
is taken into account. But to make inequality (\ref{in5}) most strong
we have to minimize $\,\overline{\Omega}(t,\Delta {\bf
R},\,\delta)\,$ with respect to $\,\delta \,$. From this point if
view, a choice when $\,1-\delta \ll 1\,$ can be preferred. At that,
expectedly, $\,\overline{\Omega}(t,\Delta {\bf R},\,\delta)\,$ is
close to above defined $\,\overline{\Omega}_{neg}(t,\Delta {\bf
R})\,$ if total pair correlation is negative.

\subsection{Finiteness of the correlation volume and
failure of the Gaussian asymptotic}

Let us consider equilibrium and hence spherically symmetric Brownian
motion what takes place in absence of the external force ($\,{\bf
f}=0\,$). Assume that at sufficiently large scales, when $\,t\gg \tau
\,$, with $\,\tau \,$ being BP's mean free-flight time, and
\[
\begin{array}{c}
\langle \Delta{\bf R}^2(t) \rangle \,=\, \int \Delta{\bf R}^2\,
V_0(t,\Delta{\bf R}\,;\,\nu_{\,0})\,d\Delta{\bf R}\,= \,6Dt\,\,\gg
3\Lambda^2\,\,\,,
\end{array}
\]
with $\,\Lambda \,$ being BP's mean free path, the probability
distribution of BP's path, $\,V_0(t,\Delta {\bf R}\,;\,\nu_{\,0})\,$,
tends to the Gaussian (\ref{vg}). Then it depends on gas density by
way of the BP's diffusivity $\,D\,$ only, and inequalities
(\ref{in4}) or (\ref{in5}) produce, together with (\ref{n2}),
\begin{eqnarray}
\fl \,\,\,\,\,\,\,\,\,\,\,\,\,\,\,\,\,\,\,\,\,\,\left
[\,c_1(t,\Delta{\bf R})\,+\,\left (\frac {\Delta{\bf R}^2}{4Dt}-\frac
32 \right )\left (\frac {\partial\, \ln D}{\partial\, \ln
\nu_{\,0}}\right )\,T\left(\frac {\partial \nu_{\,0}}{\partial
P}\right)_T\,\right ]V_G(t,\Delta{\bf R})\,\geq\,0\,\label{ga}
\end{eqnarray}
where, respectively,
\[
\fl \,\,\,\,c_1(t,\Delta{\bf R})\,=\,\nu_{\,0}\,
\max{F_1^{(eq)}}\,\,\overline{\Omega}_{neg}(t,\Delta {\bf
R})\,\,\,,\,\,\,\,\,\,c_1(t,\Delta{\bf
R})\,=\,\nu_{0}\,\max\,{F_1^{(eq)}}\,\,\overline{\Omega}(t,\Delta
{\bf R},\,\delta) \,
\]
In case of gas, certainly, BP's diffusivity is a decreasing function
of the density, $\,\partial \ln D/\partial \ln \nu_{\,0}<0\,$. Hence,
the second addend in square bracket becomes negative at
$\,z=\Delta{\bf R}^2/4Dt\,>\,3/2\,$. As the consequence, inequality
(\ref{ga}) can be satisfied if and only if at large values of $\,z\,$
quantity $\,c_1(t,\Delta{\bf R})\,$ grows at least proportionally to
$\,z\,$.

Notice that for not too dense gas (all the more, for dilute one) we
can write $\,\Lambda =v_T\tau\,$ and $\,D\approx v_T\Lambda
=\,v_T^2\tau\,$, where $\,v_T\sim \sqrt{T/M}\,$ is characteristic
thermal velocity of BP. Besides, for molecular-size BP (whose mass is
comparable with atomic mass) $\,v_T\sim v_s\,$. This makes it obvious
that limits of the quasi-uniform gas perturbation (see section 2.5)
practically allows arbitrary large values of $\,z\,$, up to $\,z\sim
v_s^2t_0/4D\sim t_0/\tau\,$, where $\,t_0\,$ is total duration of the
``full-scale experiment''. Therefore inequality (\ref{ga}) requires
from $\,c_1(t,\Delta{\bf R})\,$ and thus from
$\,\nu_{0}\overline{\Omega}\,$, with $\,\overline{\Omega}\,$ being
minimum of (\ref{av}) and (\ref{av1}), ability to achieve as large
values as $\,t_0/\tau\,$ (where $\,t_0\,$, in its turn, is arbitrary
large).

For dilute gas, more concretely, it is known that \,
$\,\max\,{F_1^{(eq)}}=1\,$,\, $\,T\,\partial \nu_{\,0}/\partial
P\,=1\,$,\, and $\,\Lambda =(\pi r_B^2\nu_{\,0} )^{-1}\propto
1/\nu_{\,0}\,$, where $\,r_B\,$ is effective radius of BP-atom
short-range repulsive interaction. Consequently,\, $\,\partial \ln
D/\partial \ln \nu_{\,0}=-1\,$, and (\ref{ga}) satisfies only when
$\,\nu_{0}\overline{\Omega}+3/2\,\geq\,z=\Delta{\bf R}^2/4Dt\,$ at
any $\,z\,$.

In fact, {\bf we are faced with dilemma: either asymptotical
statistics of Brownian path is Gaussian or volume of pair correlation
is bounded above}.

What is better? In the first variant, ``the law of large numbers''
and conventional stochastic picture of Brownian motion hold true. But
the strange enough requirement of unrestrictedly wide spatial
statistical correlations indicates presence of self-contradictions in
such theory. In the second variant, ``the law of large numbers'' does
not work. But there are no nonphysical requirements and no
contradictions.

Undoubtedly, we have to prefer this second variant and claim that
Brownian trajectory can not be imitated with the help of coin tossing
or dice tossing or other ``statistically independent'' random events.

\subsection{Volume of pair correlation (volume of collisions)}

In order to make our consideration more pointed and {\it a fortiori}
exclude from it any ``collective'' or ``hydrodynamic'' effects, a toy
named ``the Boltzmann-Grad limit'' (BGL) is very useful. In this
limit $\,\nu_{0}\rightarrow\infty\,$ while $\,r_B\sim r_A \rightarrow
0\,$ ($\,r_B\,$ and $\,r_A\,$ are radii of short-range repulsive
BP-atom and atom-atom interactions) in such way that gas non-ideality
parameters $\,4\pi r_A^3/3\,$ and $\,4\pi r_B^3/3\,$ vanish but mean
free paths of of BP, $\,\Lambda =(\pi r_B^2\nu_0 )^{-1}\,$, and
atoms, $\,\lambda =(\pi r_A^2\nu_0 )^{-1}\,$, stay fixed. At that, BP
collides with only infinitely small portion of atoms what surround
it, therefore its previous collisions in no way can influence next
ones.

Folklore of kinetics includes opinion that at least under BGL the
Boltzmann equation is true ``zero-order approximation'' for
finite-density gas kinetics, and the corresponding Boltzmann-Lorentz
equation (see e.g. \cite{re}) for molecular BP (test atom or particle
of rare impurity) is exact zero-order approximation for molecular
Brownian motion. Since this equation inevitably yields \cite{re,p1}
the Gaussian asymptotic, a best chance to resolve above formulated
dilemma in detail is to consider pair correlations in the course of
the BGL. For simplicity, again at $\,{\bf f}=0\,$.

Recall a few things about pair CF, $\,V_1(t,{\bf R}_0,{\bf r},{\bf
P}_0,{\bf p}|{\bf R}^{\prime})\,$, already known from conventional
theory \cite{re,bog,bal}. In respect to relative distance between two
particles (in our case, BP and an atom), $\,{\bf r}-{\bf R}_0\,$, the
pair correlation is accumulated inside a ``{\it collision cylinder}''
which has radius $\,\approx r_B\,$ and is directed in parallel to
relative velocity of the particles, $\,{\bf v}-{\bf V}_0={\bf
p}/m-{\bf P}_0/M\,$. Importantly, characteristic value of the pair CF
in this cylinder is comparable with product of one-particle DF. In
our case this means that magnitude of $\,V_1(t,{\bf R}_0,{\bf r},{\bf
P}_0,{\bf p}|{\bf R}^{\prime})\,$ (in particular, above defined
quantity $\,h(t,\Delta {\bf R},{\bf P}_0,{\bf p})\,$) is of order of
first right-hand term in (\ref{cf1}). As the consequence, magnitude
of pair correlation, as well as $\,V_0(t,{\bf R}_0,{\bf P}_0|{\bf
R}^{\prime})\,$, keeps safe under BGL, although inside more and more
narrow {\it collision cylinder} only.

But what is spread of pair correlation along the cylinder?
Unfortunately, conventional theory never was interested in this
issue, but in fact it reserves the spread to be infinite, at least at
$\,({\bf v}-{\bf V}_0)\cdot ({\bf r}-{\bf R}_0)>0\,$, i.e. for
particles flying away after collision. As the consequence, integral
$\,\int V_1(t,{\bf R}_0,{\bf r},{\bf P}_0,{\bf p}|{\bf
R}^{\prime})\,d{\bf r}\,$, integral in (\ref{r21}) and corresponding
correlation volumes all turn to infinity.

This became possible since the theory \cite{re,bog,bal} neglected
contributions from three-particle and other higher-order
correlations, $\,V_2\,$, $\,V_3\,$, etc., although they involve
collisions of the pair under attention with ``third particles'' (the
rest of gas). That are not literally three-particle collisions but
chains of (actual or virtual) pair collisions \cite{i1,i2,p1}. It is
not too hard to look after that magnitude of $\,n$-order CF on
corresponding sets in $\,n$-particle phase spaces at $\,n=3,...\,$,
like at $\,n=2\,$, is of order of product of $\,n\,$ one-particles DF
irrespective to BGL. Therefore, due to collisions with ``third
particles'', the pair correlation of BP with atoms disappears when
$\,|{\bf r}-{\bf R}_0|\,$ significantly exceeds $\,\Lambda\,$.
Stronger separated particles hardly are participants of forthcoming
or happening mutual collision, even if they aim precisely one to
another.

By neglecting all that, the theory unknowingly resolves the dilemma
in favor of its first variant. That is why conventional kinetics,
including the Boltzmann equation, is not true ``zero-order
approximation'' and contradicts the virial relations.

In fact, according to above remarks, a spread of the pair correlation
along collision cylinder is finite and characterized by minimum of
mean free paths $\,\Lambda\,$ and $\,\lambda\,$. Correspondingly,
effective volume of the cylinder, or volume of pair correlation,
$\,\overline{\Omega }\,$, is finite value of order of\, $\,\pi
r_B^2\Lambda=\nu_0^{-1}\,$ (assuming, for simplicity,
$\,\lambda\sim\Lambda\,$), i.e. volume displayed per one atom. Since
our reasonings are irrespective to where and when pair collisions
take place, the estimate $\,\overline{\Omega }\sim \pi
r_B^2\Lambda=\nu_0^{-1}\,$ is independent on $\,t\,$ and $\,\Delta
{\bf R}\,$ and at once is estimate of upper boundary of the
correlation volume.

Let us consider this conclusion from the point of view of virial
relations (\ref{r}), (\ref{r2}) with (\ref{clim0}), (\ref{ve}),
(\ref{r21}) and (\ref{fo}) with (\ref{fo1}).

\subsection{God does not play dice with the Boltzmann-Grad gas}

It is easy to make sure that on the way to BGL, at any fixed function
$\,\phi({\bf r})\,$, expressions (\ref{n}), (\ref{n1}) and (\ref{n2})
simplify to
\begin{eqnarray}
\frac {\nu\{{\bf r}|\phi,{\bf R}_0\}}{\nu_{0}} \rightarrow
1+\phi({\bf r})\,\,\,,\,\,\,\,\,\, \frac {\nu(\nu_{0},\phi
)}{\nu_{0}} \rightarrow 1+\phi \,\,\,,\,\,\,\,\,\,\,\,\frac
{\widetilde{\nu }_0}{\nu_{0}}\rightarrow 1 \,\,\,,\nonumber \\
\frac {\mu\{{\bf r},{\bf p}|\psi,{\bf R}_0\}}{\nu_{0}}\,\rightarrow\,
[\,1+\psi({\bf r},{\bf p})\,]\,G_m({\bf p})\,
 \label{bgl}
\end{eqnarray}

Since change of $\,\nu_{0}\,$ during BGL is compensated by change of
$\,r_B \,$ and $\,r_A \,$ to keep constant $\,\Lambda\,$ and
$\,\lambda\,$, left sides of (\ref{r}), (\ref{r2}), (\ref{ve}) and
(\ref{r21}) become invariants of $\,\nu_{0}\,$ although keeping their
dependence on $\,\phi\,$. Consequently, all coefficients in
right-hand expansions into power series of $\,\phi\,$ become
independent on $\,\nu_{0}\,$. On the left, the expansion means
application of operators\, $\,(\partial/\partial\phi )^{n}\,$ which
is equivalent to action of operators
$\,\nu_{0}^n\,[\,\partial/\partial\nu_{0}\,]^n\,$ at fixed $\,r_B \,$
and $\,r_A \,$. Then under BGL any result of this operation also is
independent on $\,\nu_{0}\,$.

What is for relations (\ref{fo}) and (\ref{fo1}), they turn to
\begin{equation}
\fl G_m({\bf p})G_M({\bf P}_0)\,\frac {\delta V\{t,{\bf
R}^{\,\prime}|\psi ,{\bf R}_0 ,{\bf P}_0\}}{\delta \mu({\bf r},{\bf
p})}\,\,e^{-{\bf f}\cdot [\,{\bf R}^{\prime}-\,{\bf R}_0]/T}
\,=\,V_1(t,{\bf R}_0,{\bf r},-{\bf P}_0,-{\bf p}|{\bf
R}^{\,\prime}\,) \label{fo11}
\end{equation}
with variational derivative taken at $\,\mu({\bf r},{\bf p})
\,=\,\nu_{0}\,G_m({\bf p})\,$. Obviously, this derivative represents
reaction of probability distribution of BP's path, $\,{\bf
R}^{\,\prime}-{\bf R}_0\,$, to lodging, at $\,t=0\,$, in point
$\,{\bf r}\,$\, one extra atom with momentum $\,{\bf p}\,$. At that,
importantly, it is presumed that this extra atom does not disturb
initial equilibrium (of the statistical ensemble), as if it was
merely marked atom.

Far enough under BGL, of course, this lodging can have an effect at
such initial relative disposition of BP and the extra atom only which
results in their direct collision. Hence, vector $\,{\bf r}-{\bf
R}_0\,$ should belong to those part of the {\it collision cylinder}
which corresponds either to particles in {\it in}-state, i.e. flying
one towards another, or to currently colliding (interacting)
particles, i.e. separated by distance $\,|{\bf r}-{\bf R}_0|\lesssim
r_B\,$.

Besides, $\,|{\bf r}-{\bf R}_0|\,$ should not be much greater than
$\,\min{(\Lambda,\lambda)}\,$. Otherwise collisions of either BP or
extra atom with ``third particles'' (other atoms) will prevent the
desired collision. This trivial notation once again, but more
strikingly, highlights that volume occupied by the pair correlation
on right-hand side of (\ref{fo11}) can be estimated (at
$\,\lambda\sim \Lambda\,$) as $\,\overline{\Omega }\sim \pi
r_B^2\Lambda =\nu_{0}^{-\,1}\,$ irrespective to the BP's path.
Indeed, on the left in (\ref{fo11}), in view of the causality
principle, a whole future path of BP can not influence possibility of
its collision with concrete marked atom at very beginning of the
path. Moreover, the mentioned path and collision can not be mutually
statistically correlated since the marked atom and BP deal with
different non-intercrossing collections of ``third particles''.
Therefore, any dependence of the right-hand pair CF on $\,t\,$ and
$\,\Delta {\bf R}\,$ characterizes its magnitude but not its spread.

Now we can say almost with confidence, that correlation volumes
$\,\overline{\Omega }_{neg}(t,\Delta {\bf R})\,$, $\,\Omega(t,\Delta
{\bf R},{\bf P}_0,{\bf p},\delta)=\Omega(\delta )\,$ and
$\,\overline{\Omega }(t,\Delta {\bf R},\delta)\,$, introduced in
sections 3.1 and 3.2, have upper boundary $\,\sim \nu_{0}^{-\,1}\,$
independent on $\,t\,$ and $\,\Delta {\bf R}\,$ or $\,z\,$,\, and
should be written simply as $\,\overline{\Omega }_{neg}\,$,
$\,\Omega({\bf P}_0,{\bf p},\delta)=\Omega(\delta )\,$ and
$\,\overline{\Omega }(\delta)\,$. Correspondingly,
$\,c_1(t,\Delta{\bf R})\,$ in (\ref{ga}) is not a function but a
constant of order of unit.

Thus, the second resolution of the dilemma from section 3.3 is
acceptable only. We can say that the god of mechanics does not play
dice. He disposes of stronger means: Hamiltonian dynamics of many
particles ($\,N>t_0/\tau \,$, according to \cite{pro,lp,p1}) is able
to create much more rich randomness than dice tossing can do.

\subsection{Probability distribution of BP's path
possesses power-law long tails}

Now we are ready to discuss what kind of asymptotic of BP's path
distribution is really allowed instead of the Gaussian one.

According to previous section, let us rewrite inequalities
(\ref{in4}) or (\ref{in5}) in the form
\begin{eqnarray}
c_1\,V_0(t,\Delta{\bf R}\,;\,\nu_{\,0})+\,\widetilde{\nu}_0 \,\frac
{\partial\, V_0(t,\Delta{\bf R}\,;\,\nu_{\,0})}{\partial\,\nu_{\,0}}
\,\geq\,0\,\,,\label{nga}
\end{eqnarray}
Here $\,c_1\,$ is a constant whose upper estimates obtained in two
different ways look as
\[
c_1\,=\,\nu_{\,0}\,
\max{F_1^{(eq)}}\,\,\overline{\Omega}_{neg}\,\,\,,\,\,\,\,\,\,
c_1\,=\,\nu_{0}\,\max\,{F_1^{(eq)}}\,\,\overline{\Omega}(\delta) \,
\]
with $\,\overline{\Omega}_{neg}\,$ and $\,\overline{\Omega}(\delta)
\,$ being volumes of pair correlation (volumes of collision) defined
by (\ref{av}) and (\ref{av1}) and bounded above by a value
$\,\sim\nu_{\,0}^{-1}\,$\, (volume per one atom), so that $\,c_1\sim
1\,$. Probably, there are methods to estimate $\,c_1\,$ differently
from sections 3.1 and 3.2. Anyway, to make inequality (\ref{nga})
stronger, we should take minimum of all available estimates.

For simplicity, consider equilibrium Brownian motion at $\,{\bf
f}=0\,$. In case of dilute enough gas, or even liquid in three or
more dimensions, when hydrodynamical contributions to BP's velocity
are far from domination, and the ``diffusion law'' $\,\Delta{\bf
R}^2\propto t\,$\, holds, it seems natural if asymptotically, at
$\,t\gg \tau \,$, the BP's path distribution $\,V_0(t,\Delta {\bf
R}\,;\,\nu_{\,0})\,$ is characterized, similarly to $\,V_G(t,\Delta
{\bf R})\,$, by a single parameter, that is diffusivity. For 3-D
space,
\begin{equation}
V_0(t,\Delta{\bf R}\,;\,\nu_{\,0}) \rightarrow
(4Dt)^{-3/2}\,\Psi(\Delta {\bf R}^2/4Dt)\,\label{as0}
\end{equation}
Here it is presumed that
\begin{equation}
\int \Psi({\bf a}^2)\,d{\bf a}\,=1\,\,\,,\,\,\,\,\,\,\int {\bf
a}^2\Psi({\bf a}^2)\,d{\bf a}\,=3/2\,\label{pres}
\end{equation}
The first of these requirements is the normalization condition, while
the second means that\, $\,\langle \Delta{\bf R}^2(t) \rangle
=6Dt\,$\, and in fact serves as quantitative definition of the
diffusivity. Then, inequality (\ref{nga}) yields
\begin{equation}
\alpha\,\Psi(z) +z\,\frac {d\,\Psi(z)}{d\,z} \,\geq
\,0\,\,\,\,\,,\,\,\,\,\,\,\,\,\alpha \,\equiv\,\frac 32\,+\,c_1\left
|\,\frac {\widetilde{\nu}_0 }{D}\,\frac {\partial
D}{\partial\nu_0}\right |^{-\,1} \,\,,\label{as1}
\end{equation}
if diffusivity is a decreasing function of density,\, $\,\partial
D/\partial\nu_0\,<0\,$,\, and
\[
\alpha\,\Psi(z) +z\,\frac {d\,\Psi(z)}{d\,z} \,\leq
\,0\,\,\,\,\,,\,\,\,\,\,\,\,\,\alpha \,\equiv\,\frac 32\,-\,c_1\left
|\,\frac {\widetilde{\nu}_0 }{D}\,\frac {\partial
D}{\partial\nu_0}\right |^{-\,1} \,
\]
in opposite case\, $\,\partial D/\partial\nu_0\,<0\,$\,.

Confining ourselves by the first case, we see that function
$\,\Psi(z)\,$ must have power-law long tail: \, $\,\Psi(z)\propto\,
z^{-\,\beta }\,$\, at $\,z\rightarrow\infty\,$\,, where $\,\beta\leq
\alpha\,$\,. At that, in order to satisfy second of conditions
(\ref{pres}), exponent\, $\,\alpha \,$ should be greater than
$\,5/2\,$.

Example of such behavior, with maximally possible $\,\beta
=\alpha\,$, is presented by
\begin{equation}
\fl \,\,\,\,\,\,\,\,\,\,\,\,\,\,\,\Psi(z)\,=\,\frac {\gamma(\alpha
)}{(1+z)^{\,\alpha }}\,\,\,\,\,\,\,,\,\,\,\,\,V_0(t,\Delta{\bf
R}\,;\,\nu_{\,0}) \rightarrow\,\frac {\gamma(\alpha )}{(4D^{\prime
\,}t)^{3/2}}\left (1+\frac {\Delta{\bf R}^2}{4D^{\prime \,}t}\right
)^{-\,\alpha } \,,\label{ex}
\end{equation}
where\,\, $\,\gamma(\alpha )\,=\,\pi^{-3/2}\,\Gamma(\alpha
)/\Gamma(\alpha -3/2)\,$\, and\, $\,D^{\prime \,}=(\alpha-5/2)D\,$.

Unpleasant aspect of such behavior is unboundedness of high enough
statistical moments of $\,\Delta{\bf R}\,$. However, recall that real
Brownian motion in addition to diffusivity has at least one more
important parameter, namely, the BP's thermal velocity $\,v_T\,$.
Then $\,V_0(t,\Delta{\bf R})\,$ should be considered as a function of
two dimensionless parameters, $\,z=\Delta{\bf R}^2/4Dt\,$ and
$\,y={\bf R}^2/v_T^2t^2\,$, and asymptotically instead of (\ref{as1})
we can write
\begin{equation}
\fl V_0(t,\Delta {\bf R})\rightarrow \frac {1}{(4Dt)^{3/2}}\,
\Phi\left (\frac {\Delta{\bf R}^2}{4Dt}\,,\frac {\Delta {\bf
R}^2}{v_T^2\, t^2}\right )\rightarrow \frac
{1}{(4Dt)^{3/2}}\,\Psi\left (\frac {\Delta{\bf
R}^2}{4Dt}\right)\Theta \left (\frac {\Delta {\bf
R}^2}{v_T^2\,t^2}\right )\,\,\label{as2}
\end{equation}
Here $\,\Psi(z)\,$ as before satisfies (\ref{pres}), while function
$\,\Theta(y)\approx 1\,$ at $\,y\ll 1\,$ and fast enough tends to
zero at $\,y>1\,$ thus sharply cutting $\,V_0(t,\Delta{\bf R})\,$
when\, $\,|{\bf R}|>v_Tt\,$\,. This variant also is allowed by
inequality (\ref{nga}), because $\,v_T\,$ is independent on density.
And now all statistical moments of BP's are finite.

Nevertheless, from the point of view of higher-order moments, such
asymptotic still is not exhaustive, excepting case of dilute gas. For
sufficiently dense gas (all the more, liquid) we should take into
account, at the minimum, such third (density-dependent) parameter as
isothermal speed of sound,
$\,v_s=\sqrt{T\nu_0/m\widetilde{\nu}_0}\,$, with
$\,\widetilde{\nu}_0\,$ presented by (\ref{n2}). At the same time,
such complications practically do not change main probabilistic
characteristics of BP's path, as well as its mean square behavior,
and in no case cancel the long power-law tails.

\subsection{Comparison with solutions of the BBGKY equations}

Previous results, obtained from the first-order virial relations
(one- and two-particle CF) only, are in remarkable qualitative (and
even semi-quantitative) agreement with results obtained in \cite{p1}
from infinite chain of roughened BBGKY equations describing a test
(marked) gas atom in the role of BP in the Boltzmann-Grad gas. But,
of course, our approach here could not give numeric value of exponent
$\,\alpha\,$ in (\ref{as1}) and (\ref{ex}). Approximate analysis in
\cite{p1} gave distribution (\ref{ngl}), that is $\,\alpha =7/2\,$,
which corresponds to $\,c_1=\nu_0\overline{\Omega }=2\,$.

More precisely, solution obtained in \cite{p1} indeed has the form
(\ref{as2}). At that, the cut-off function $\,\Theta\,$ works already
at fourth-order statistical moment, yielding
\begin{equation}
\langle\, \Delta{\bf R}^4(t) \,\rangle \,=\,\int \Delta{\bf
R}^4\,V_0(t,\Delta {\bf R})\,d\Delta{\bf
R}\,\rightarrow\,3\,\langle\, \Delta{\bf R}^2(t) \,\rangle
^2\,\ln{\frac {t}{\tau }} \,\label{fm}
\end{equation}
with $\,\langle \Delta{\bf R}^2(t) \rangle =6Dt\,$. This result
looks\, as\, if\, BP's diffusivity was fluctuating quantity,
$\,\widetilde{D}\,$, with mean $\,\langle \widetilde{D} \rangle =D\,$
and variance $\,\langle \widetilde{D}^2 \rangle -\langle
\widetilde{D} \rangle^2 \approx D^2\,\ln{(t/\tau )}\,$ which depends
on total duration of BP observation, $\,t\,$.

From the point of view of (\ref{ngl}) variance of the fluctuating
diffusivity, $\,\widetilde{D}\,$, is infinite. But probability
distribution of $\,\widetilde{D}\,$ is quite certain. One uncovers it
representing (\ref{ngl}) as superposition of Gaussians with various
values of diffusivity:
\begin{eqnarray}
\fl V_0(t,\Delta{\bf R}) \rightarrow\frac {\Gamma(7/2)}{(4\pi
Dt)^{3/2}}\left [1+\frac {\Delta{\bf R}^2}{4Dt}\right ]^{-\,7/2
}=\int \frac {1}{(4\pi \zeta Dt)^{3/2}}\,\exp{\left [-\frac
{\Delta{\bf R}^2}{4\zeta Dt}\right
]}\,w(\zeta)\,d\zeta\,\,\,,\nonumber \\
\,\,\,\,\,\,\,\,\,\,w(\zeta)\,=\,\frac {1}{\zeta ^3}\,\exp{\left
(-\frac {1}{\zeta }\right)}\,\,\,,\label{sup}
\end{eqnarray}
where\, $\,\zeta\,$ represents $\,\widetilde{D}/D\,$ while
$\,w(\zeta)\,$ is probability density of $\,\zeta\,$.

\begin{figure}
\includegraphics[width=13cm]{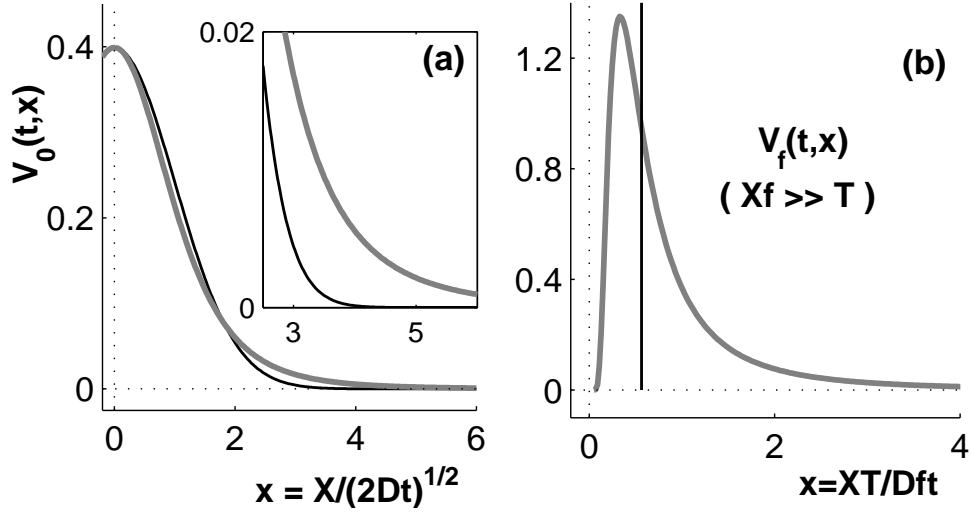}
\caption{\label{fig1} (a)\, Projections onto $\,X$-axis of reduced
Gaussian distribution (\ref{vg}), $\,(2\pi )^{-1/2}\exp{(-x^2/2)}\,$
(thin black curve), and reduced rescaled non-Gaussian distribution
(\ref{ngl}), $\,(2\pi )^{-1/2}(1+8x^2/9\pi )^{-5/2}\,$ (thick gray
curve); the inset magnifies tails of the distributions.\, (b)\,
Reduced distribution (\ref{x}), $\,x^{-3}\exp{(-1/x)}\,$ (thick gray
curve), of drift displacement in comparison with reduced rescaled
result, $\,\delta(x-16/9\pi )\,$ (black vertical line), of usual
theory.}
\end{figure}

At one and the same $\,D\,$ distribution (\ref{ngl}) or (\ref{sup})
seems rather dissimilar to the Gaussian (\ref{vg}). But in fact the
only essential difference between them is long tail of (\ref{ngl}).
To make this visible, we may rescale (\ref{ngl}) by suitable increase
(``renormalization'') of its diffusivity and compare one-dimensional
projections of (\ref{vg}) and (\ref{ngl}) onto some axis $\,X\,$.
Left plot on Figure 1 illustrates such comparison, with rescaling
factor $\,3\sqrt{\pi}/4\,$ which equalizes heights of both
distributions at $\,X=0\,$.

Although in \cite{p1} only thermodynamically equilibrium Brownian
motion was considered, the integral representation (\ref{sup}), as
combined with FDR (\ref{r0}), can be used to predict statistics of
non-equilibrium Brownian motion at $\,{\bf f}\neq 0\,$.
Reasonableness of such operation, for a weak non-equilibrium, was
confirmed e.g. in \cite{bk2,pr}. The word ``weak'' means that BP's
drift velocity $\,{\bf v}_d\,$ is much smaller than $\,v_T\,$ and
$\,v_s\,$ and therefore is linear function of the external force:
\[
\begin{array}{c}
\fl \,\,\,\,\,\,\,\,\,\,\,\,\,\,\,\,\,\,\,\,\,\,\,\,\langle \Delta
{\bf R}(t)\rangle\,=\,\int \Delta {\bf R}\,\,V_0(t,\Delta {\bf
R})\,d\Delta {\bf R}\,=\,{\bf v}_d t\,\,\,\,,\,\,\,\,\,\,\,{\bf
v}_d\,=\,D{\bf f}/T\,
\end{array}
\]
Here $\,D/T\,$ is mobility of BP, in agreement with the Einstein
relation \cite{ae1,ae2} which, by the way, directly follows from
(\ref{r00}) or (\ref{r0}) (see also \cite{i2,p,bk3}).

To underline dependence of $\,V_0(t,\Delta{\bf R})\,$ on the external
force, let us denote it as $\,V_f(t,\Delta{\bf R})\,$. Then, if we do
not wish to look on so far distances as $\,v_Tt\,$ (all the more,
$\,|{\bf f}|t^2/2M\,$), we can find asymptotic of $\,V_f(t,\Delta{\bf
R})\,$ at $\,t\gg \tau\,$ in the form (\ref{sup}) but replacing
equilibrium Gaussian components by non-equilibrium ones:
\begin{eqnarray}
\fl \,\,\,\,\,\,\,\,\,\,\,\,\,\,\,\,\,\,\,\,\,\,\,V_f(t,\Delta{\bf
R}) \rightarrow \int \frac {1}{(4\pi \zeta Dt)^{3/2}}\,\exp{\left
[-\frac {\left (\Delta{\bf R}-\zeta D{\bf f}\,t/T\,\right )^2}{4\zeta
Dt}\right ]}\,w(\zeta)\,d\zeta\,\label{neq}
\end{eqnarray}
Recollecting that $\,\partial \ln{D}/\partial \ln{\nu_0}=-1\,$ under
BGL, it is easy to make sure that this expression satisfies
inequality (\ref{nga}), that is
\[
c_1\,V_f(t,\Delta{\bf R})-\,D\,\frac {\partial\, V_f(t,\Delta{\bf
R})}{\partial\,D}
\,\geq\,0\,\,\,\,,\,\,\,\,\,\,\,\,\,\,c_1\,=\,2\,\,\,,
\]
as well as equality (\ref{r0}). In short, (\ref{neq}) agrees with
both virial relations and FDR.

Consider (\ref{neq}) at a time when drift of BP exceeds its
diffusion, $\,|{\bf v}_d|t >\sqrt{6Dt}\,$. In other words, when
$\,{\bf f}\cdot{\bf v}_d\,t\gg T\,$, i.e. work of the force much
exceeds thermal energy per degree of freedom. At that, distribution
$\,V_f(t,\Delta{\bf R})\,$ becomes highly asymmetric and anisotropic.
Therefore we have to distinguish BP's displacement $\,X\,$ along
direction of the force $\,{\bf f}\,$ and two orthogonal displacements
$\,Y\,$ and $\,Z\,$. In such notations $\,\Delta{\bf R}=\{X,Y,Z\}\,$,
and (\ref{neq}) yields
\begin{eqnarray}
\fl \,\,\,\,\,\,\,\,\,\,\,\,\,\,\,\,\,V_f(t,\{X,Y,Z\})\,
\rightarrow\, \frac {|{\bf f}|}{4\pi TX}\,\exp{\left [-\frac {|{\bf
f}|(Y^2+Z^2)}{4TX}\right ]}\,\cdot\,\frac {{\bf v}_d^2
t^2}{X^3}\,\exp{\left (-\frac {|{\bf v}_d|t}{X}\right )}
\,\label{lneq}
\end{eqnarray}
At any fixed $\,Y\,$ and $\,Z\,$ this expression has long tail in
positive $\,X$-direction, so that
\begin{eqnarray}
\fl \,\,\,\,\,\,\,\,\,\,\,\,\,\,\,\,\,\,\,\,\,\,\,\,
\,V_f(t,X)\,\equiv \int\int V_f(t,\{X,Y,Z\})\,dY\,dZ \,=\, \frac
{{\bf v}_d^2 t^2}{X^3}\,\exp{\left (-\frac {|{\bf v}_d|t}{X}\right )}
\,\,\,,\label{x}
\end{eqnarray}
but in strictly this direction only. Right-hand part of Figure 1
shows this asymptotical distribution of BP's drift, in comparison
with drift distribution in conventional kinetics, which is nothing
but delta-function $\,\delta (X-|{\bf v}_d|t)\,$. At the same time,
\[
\begin{array}{c}
\fl \int V_f(t,\{X,Y,Z\})\,dX\, =\,\int V_0(t,\{X,Y,Z\})\,dX\,
=\,(2\pi Dt)^{-1}\,[\,1+(Y^2+Z^2)/4Dt\,]^{-3} \,
\end{array}
\]
has long tail in any direction in $\,YZ$-plane, and anybody who keeps
under observation $\,Y\,$ and $\,Z\,$ only sees no signs of BP's
drift along $\,X$-axis.

Basic idea of \cite{p1}, earlier suggested in \cite{i1,i2}), was to
consider DF averaged over ``collision boxes''. The latter are just
sets in $\,n$-particle phase spaces already mentioned in section 3.4.
For $\,n=2\,$, ``collision box'' is nothing but above exploited
``collision cylinder''. In general, ``collision box'' is a
``skeleton'' set of $\,n$-particle configurations snapped up from
chains of $\,n-1\,$ connected (actual or virtual) collisions, under
given input momenta. Thorough description of all these sets hardly is
possible, but hardly it is obligatory. Much more important thing is
that even under rough description one immediately discovers that the
Boltzmann-Grad gas is true\, {\it terra incognita}\,.

\subsection{Origin of the historical correlations and long tails
and 1/f noise}

Let us return to identity (\ref{fo11}). One and the same vector
$\,{\bf r}-{\bf R}_0\,$ on its opposite sides belongs to opposite
half of the {\it collision cylinder}. Hence, pair correlations on
right-hand side are concentrated at {\it out}-states (i.e. concern
particles flying away one from another) and at central region of the
cylinder (i.e. concern also currently interacting particles, with
$\,|{\bf r}-{\bf R}_0|\lesssim r_B\,$), while {\it in}-states are
uncorrelated.

This consequence of (\ref{fo11}) is in agreement with usual theory
\cite{bog,uf,re,bal}. The latter considers it as a good reason to
believe in ``statistical independency'' of colliding particles
(Boltzmann's ``molecular chaos''). In present theory, however,
actually colliding particles occupy central part of collision
cylinder where, according to (\ref{fo11}), $\,V_1(t,{\bf R}_0,{\bf
r},{\bf P}_0,{\bf p}|{\bf R}^{\,\prime})\,$ is non-zero, and thus
these particles are mutually correlated and statistically dependent.
Hence, statistical dependency is not ``exported from outside'' but
occurs at the time of collision.

The point is that $\,V_1(t,{\bf R}_0,{\bf r},{\bf P}_0,{\bf p}|{\bf
R}^{\,\prime})\,$ (as well as higher-order CF) represents
``historical correlations'' (see section 2.3): it treats current
collision as not a separate event but last term of long random
sequence of BP's collisions with atoms on its way from $\,{\bf
R}^{\,\prime}\,$ to $\,{\bf R}_0\,$. Therefore statistical
dependencies involved by $\,V_1(t,{\bf R}_0,{\bf r},{\bf P}_0,{\bf
p}|{\bf R}^{\,\prime})\,$ should be addressed not to currently
colliding particles (BP and atom) themselves but to long-range
fluctuations in relative frequency of BP's collisions.

With the purpose to make certain that this is only reasonable
understanding of the correlations, in particular spatial ones,
$\,V_n(t,\Delta {\bf R}\,;\,\nu_{\,0}\,)\,$ ($\,n\geq 1\,$), let us
introduce, in terms of section 3.2, function
\begin{eqnarray}
\fl \,\,\,\,\,\,\,\,\,\,\,\,\,\,\,\,W_1(t,\Delta {\bf
R})\,=\,\frac {\nu_{\,0}}{1-\delta}\int\int\left[\int_{\Omega({\bf
P}_0,\,{\bf p},\,\delta\,)} F_1(t,{\bf R}_0,{\bf r},{\bf P}_0,{\bf
p}|{\bf R}^{\prime})\,d{\bf r}\right ]d{\bf p}\,d{\bf P}_0\,\nonumber
\end{eqnarray}
which is direct analogue of function $\,W_2\,$ from \cite{p1} and
satisfies
\[
\begin{array}{c}
\,\,\,\,\,\,\int W_1(t,\Delta {\bf R})\,d\Delta {\bf
R}\,=\,c_1=\,\nu_{\,0}\,\overline{\Omega}(\delta)\,
\end{array}
\]
Evidently, $\,W_1(t,\Delta {\bf R})\,$ is probability density of
finding BP at point $\,\Delta {\bf R}\,$ (after start from coordinate
origin) and simultaneously some atom in its close vicinity in some
{\it in}-state or {\it out}-state of their mutual collision.
Consequently, ratio
\[
\begin{array}{c}
W_1(t,\Delta {\bf R})/V_0(t,\Delta {\bf
R})\,=\,\mathcal{P}_{post}(t,\Delta {\bf R})
\end{array}
\]
presents a measure of conditional probability of BP's collision under
condition that path $\,\Delta {\bf R}\,$ is known. The subscript
``\,{\it post}\,'' underlines that in essence this is {\it a
posteriori} probability. In absence of historical correlations
between BP and atoms we would have $\,W_1(t,\Delta {\bf R})=c_1
V_0(t,\Delta {\bf R})\,$ and above ratio would reduce to
unconditional {\it a\,\,priori}\, probability measure:
$\,\mathcal{P}_{prior}=\nu_{\,0}\,\overline{\Omega}(\delta)\,$.

In conventional kinetics, based on concepts like ``probability of
collision'', any difference between $\,\mathcal{P}_{post}\,$ and
$\,\mathcal{P}_{prior}\,$ is unthinkable since it means dependence of
``probability of collision'' on wherefrom BP had started in former
times. Therefore factual difference of $\,\mathcal{P}_{post}\,$ from
$\,\mathcal{P}_{prior}\,$, indicated by the virial relations, says
that a random point process formed by BP' collisions in not a usual
Poisson process. Then, what is it? The answer is rather obvious: it
is Poisson process but supplied with such (relatively slow) scaleless
random variations of ``probability of collision'' which exclude
possibility to get it by time averaging. Indeed, on second thought it
is clear that Poissonian statistics introduces ``too strict
disorder'' in collisions to be likely.

Such interpretation of inter-particle correlations which keep safe
even under BGL was expounded in \cite{i1,i2} (besides, it was in part
developed in \cite{pro,lp,p71,p1}, and its most principal aspects
were anticipated already in \cite{bk1,bk2,bk3,pr}). It was
demonstrated that fluctuations in ``probability of collision'' (or,
more precisely, relative frequency of collisions) possess scaleless
1/f-type spectrum (i.e. represent ``1/f-noise'') and may be described
also as 1/f fluctuations of BP's diffusivity (and mobility). Now we
arrived at principally same results after start from virial
relations.

Scaleless character of the fluctuations is due to fact that the
system under consideration constantly forgets history of BP's
collisions and therefore has neither stimulus nor means to force
relative frequency of collisions to be certain. To keep it certain at
arbitrary large time scales, the system, in opposite, should keep in
mind arbitrary old history. In this sense, {\it a priori}\,
equalization of $\,\mathcal{P}_{post}\,$ and
$\,\mathcal{P}_{prior}\,$ acts as infinitely long memory. Notice that
in comparison with Poissonian statistics of collisions a real one to
some extent resembles Bose statistics.

Now let us compare $\,\mathcal{P}_{post}\,$ and
$\,\mathcal{P}_{prior}\,$. From section 3.5 it follows that sign of
$\,V_1(t,{\bf R}_0,{\bf r},{\bf P}_0,{\bf p}|{\bf R}^{\,\prime})\,$
is determined by $\,t\,$ and $\,\Delta {\bf R}\,$ only. Taking this
into account, in the spirit of section 3.2 we can derive one more
inequality,
\begin{eqnarray}
\fl \,\,\,\,\,\,\,\,\,\,\,\,\,\,\,\,\,\,\,\,\,\,\,\,\,\,\left |\,
W_1(t,\Delta {\bf R})\,-\,c_1\,V_0(t,\Delta {\bf R})\,-\frac
{V_1(t,\Delta {\bf R})}{1-\delta }\,\right |\,\leq\, \delta \,\left
|\,\frac {V_1(t,\Delta {\bf R})}{1-\delta }\,\right |\,\label{lin}
\end{eqnarray}
Being combined with (\ref{r21}) it implies that
\begin{eqnarray}
\frac {\mathcal{P}_{post}(t,\Delta {\bf
R})}{\mathcal{P}_{prior}}\,\,\leq\,\,1\,+\,\frac {1}{c_1}\,\frac
{\partial\, \ln V_0(t,\Delta {\bf R})}{\partial\,
\ln\nu_{\,0}}\,<\,1\,\label{sin}
\end{eqnarray}
when\, $\,\partial V_0(t,\Delta {\bf R})/\partial \nu_{0}\,<\,0\,$\,.
When, in opposite,\, $\,\partial V_0(t,\Delta {\bf R})/\partial
\nu_{0}\,>\,0\,$\,, then both inequality signs in (\ref{sin}) should
be inverted.

Inequality (\ref{sin}) represents strengthened form of (\ref{r21}).
At tails of $\,V_0(t,\Delta {\bf R})\,$ always $\,\partial
V_0(t,\Delta {\bf R})/\partial \nu_{0}\,<\,0\,$, and it shows that,
naturally, {\it a posteriori} probability of BP's collisions far
enough at tails is smaller than {\it a priori} one.

Notice that if $\,c_1\,$ was an exact value, determined by minimum of
all possible estimates of pair correlation volume (see section 3.6
and above), then middle expression in (\ref{sin}) would achieve its
theoretical minimum too. Hence, in fact left inequality should be
replaced by equality. Similar conclusion is valid also in respect to
its alternative at $\,\partial V_0(t,\Delta {\bf R})/\partial
\nu_{0}\,>\,0\,$. Thus we can write
\begin{eqnarray}
\fl\,\,\,\,\,\,\,\,\,\,\,\,\,\,\,\,\,\,\,\,\,\,\, \frac
{\mathcal{P}_{post}(t,\Delta {\bf R})}{\mathcal{P}_{prior}}\,\,=
\,1\,+\,\frac {1}{c_1}\,\frac {\partial\, \ln V_0(t,\Delta {\bf
R})}{\partial\, \ln\nu_{\,0}}\,=\,1\,-\,\frac {1}{c_1}\,\frac
{\partial\, \ln V_0(t,\Delta {\bf R})}{\partial\,\ln D
}\,\label{tail0}
\end{eqnarray}
The last equality here relates to the Boltzmann-Grad gas. For
distribution (\ref{ngl}) or (\ref{sup}),
\begin{eqnarray}
\frac {\mathcal{P}_{post}(t,\Delta {\bf
R})}{\mathcal{P}_{prior}}\,\rightarrow\, \frac 74 \left (1+\frac
{\Delta {\bf R}^2}{4Dt}\right )^{-\,1}\,\label{rel}
\end{eqnarray}
So substantial changeability of the {\it a posteriori} ``probability
of collision'' shows that in fact this quantity has no certain value
obtainable by time averaging.

\section{Conclusion}

Being guided by wrong idea of ``statistical independency'' of
colliding particles (Boltzmann's ``molecular chaos''), classical gas
kinetics neglected statistical inter-particle correlations which
inevitably arise in spatially non-uniform Gibbsian statistical
ensembles. Thus the concept of {\it a priori}\, definable
``probability of collision'' was unknowingly imposed on the theory as
characteristics of a concrete particle trajectory. As for Brownian
motion (self-diffusion) of gas particles, the result was ``the law of
large numbers'' stating that probability distribution $\,V_0(t,\Delta
{\bf R})\,$ of path of molecular Brownian particle (BP) is drawn
towards the Gaussian distribution.

We derived exact virial expansion connecting response of
$\,V_0(t,\Delta {\bf R})\,$ to gas perturbations, from one hand, and
pair and many-particle ``historical'' statistical correlations
between the BP's path and gas, from the other hand. Specificity of
``historical'' correlations is that they are just products of initial
spatial non-uniformity of an ensemble. Thus existence of such
correlations is proved. At the same time, we showed that finiteness
of spatial spread of these correlations is incompatible with Gaussian
asymptotic of $\,V_0(t,\Delta {\bf R})\,$ even in dilute gas (under
the Boltzmann-Grad limit). Then we demonstrated that the spread is
really finite and, consequently, $\,V_0(t,\Delta {\bf R})\,$ has
essentially non-Gaussian form, with power-law long tails (lasting up
$\,|\Delta {\bf R}|\sim v_Tt\,$). These results mean (i) that BP's
path can not be divided into ``statistically independent'' events or
pieces, and, as combined with our previously published
\cite{i1,bk1,bk2,bk3} and unpublished \cite{i2,pro,lp,p71,p1}
results, (ii) that ``probability of collision'' or, equivalently,
diffusivity (and mobility) of BP in fact undergo scaleless
fluctuations like the so-called 1/f-noise.

In principle, these conclusions extend to the kinetics as the whole.
If applied to charge carriers in semiconductors, they naturally
explain the ``inherent 1/f-noise'' under experimental investigation
during many years \cite{hkv,tv}. However, one has to revise kinetics
of electron-phonon systems (as well as phonon systems themselves
\cite{i3}), again removing from it {\it a priori}\, ``probability of
collision'' and ``statistical independencies'', and returning to
honest analysis of infinite chains of many-particle statistical
correlations. Perhaps, a time of such revision is not far off.

I am grateful to Dr.\,\,Yu.\,Medvedev and Dr.\,\,I.\,Krasnyuk for
useful discussions.

\end{document}